\documentclass[a4paper,11pt]{article}
\pdfoutput=0 % if your are submitting a pdflatex (i.e. if you have
\usepackage{jheppub} % for details on the use of the package, please
\usepackage{graphicx}

\usepackage[T1]{fontenc} % if needed

\newcommand{\sss}[1]{\scriptscriptstyle #1}
\newcommand{\Ref}[1]{(\ref{#1})}
\newcommand{\be}{\begin{eqnarray}}
\newcommand{\ee}{\end{eqnarray}}
\title{\boldmath
Testing the bootstrap
constraints in the strange sector
}

\author[a,b]{Kirill M.~Semenov-Tian-Shansky, %\note{Corresponding author.}
}
\author[b]{Vladimir V.~Vereshagin}

\affiliation[a]{IFPA, d\'{e}partement AGO,  Universit\'{e} de  Li\`{e}ge, \\ 4000 Li\`{e}ge,  Belgium}
\affiliation[b]{St.-Petersburg State University,  \\ St.-Petersburg,  198504, Petrodvoretz,   Russia }

\emailAdd{ksemenov@ulg.ac.be}
\emailAdd{vvv@av2467.spb.edu}

\abstract{
In this paper the bootstrap conditions that follow from the general
postulates of effective scattering theory (EST)
are checked in the strange sector. We construct the system of tree
level bootstrap constraints for the renormalization prescriptions
fixing the physical content of the theory. Then we perform the
numerical testing of corresponding sum rules for the parameters of
strange resonances. It is shown that, generally, the bootstrap
constraints turn out consistent with presently known data on the
strange resonance parameters. At the same time we point out few sum
rules which cannot be saturated with modern data and discuss the
possible reasons for such discrepancies.
}

\begin{document}
\maketitle
\flushbottom

%%%%%%%%%%%%%%%%%%%%%%%%%%%%%%%%%%%%%%%%%%%%%%%%%%%%%%%%%%%%%%%%%%%%70

\section{Introduction}

In
\cite{VVV}--\cite{KSAVVV2}
it has been made an attempt to develop the formalism necessary to
handle the infinite-component effective theories. A special point of
that formalism is that it is destined solely for the description of
scattering processes. In fact we are constructing not an effective
field theory but rather an
{\em effective scattering theory} (EST).
The key to the solution of specific problems which emerge when one
considers an infinite component effective theory is provided by the
requirement of existing the rigorously defined terms (each given by an infinite series) of Dyson perturbation expansion
at every fixed loop order. The constructive form of this requirement
results in a set of non-trivial relations for the renormalization
prescriptions (RPs) that fix the physical content of the theory -- so
called
{\em bootstrap constrains}.

In principle, once solved the system of bootstrap constraints would
give the answer to the question:
``How many independent RPs are needed to fix completely the physical
content of an infinite component EST?''and would show its true
predictive power. Unfortunately, actually we are unable to solve this
system explicitly. However, the numerical tests of sum rules for the
parameters of resonances derived from the bootstrap system can be used to check
the consistency of EST approach.

In this paper we continue the work started in
\cite{VVV}--\cite{KSAVVV2}.
As shown in
\cite{KSAVVV3},
the data on resonances in three channels of the elastic pion-nucleon
scattering process turn out to be in nice agreement with corresponding
bootstrap constraints. This gives us a hope that the main principles
forming the basis for extended perturbation scheme mirror correctly
the regularities of hadron spectrum. Therefore, it seems natural to
check if the method of
\cite{KSAVVV3}
leads to reasonable results in the case of strange resonances that
appear in elastic kaon-nucleon scattering. This process presents
special interest because, first, it is relatively well studied
experimentally and, second, the resonance spectra in
$s$-
and
$u$-channels
differ from one another. The latter circumstance provides a
possibility to exploit a far more rich system of bootstrap constraints
as compared to that used in our analysis of pion-nucleon elastic
scattering.

The main principles of constructing the extended perturbation scheme
are discussed in
\cite{VVV}--\cite{KSAVVV2}.
A detailed step-by-step instruction on their application for the case
of binary processes at tree level is presented in
\cite{KSAVVV3}.
Therefore, in this paper we refer the reader to quoted above articles for the
details of our approach and focus only on the specific points of the
particular case under consideration.

We would like to emphasize that the main goal of this paper is the analysis of the
bootstrap relations for the parameters of strange
resonances and checking their consistency with
well established experimental numbers. We do not aim to
give a precise description of the relevant amplitudes in any
kinematical domain (say, near threshold or in the resonance region).

The paper is organized similarly to
\cite{KSAVVV3}.
In Section~\ref{sec-preliminaries}
the basic formulae needed to construct the Cauchy forms for tree-level
invariant amplitudes of
$KN$
elastic scattering are presented.
Next, in
Section~\ref{sec-Cforms}
we construct the Cauchy forms in three mutually intersecting
hyperlayers. With these expressions in hand, in
Section~\ref{sec-strboot}
we derive the system of bootstrap constraints (sum rules) for the
set of renormalization prescriptions fixing the physical content of
the effective theory. Bootstrap
system restricts the values of resonance mass parameters and the
minimal (resultant) triple coupling constants%
\footnote{We refer the reader to
Ref.~\cite{KSAVVV2}
for the detailed explanation of the terminology.}.
The results of numerical testing of the corresponding sum rules for
spectrum parameters are discussed in
Section~\ref{S_NumTests}.
We present a large set of sum rules that are well saturated by the known
data. Next we consider certain sum rules which are not saturated by
the parameters of presently known resonances and discuss the possible
sources of discrepancies. The conclusions are given in
Section~\ref{conclusion}.

%%%%%%%%%%%%%%%%%%%%%%%%%%%%%%%%%%%%%%%%%%%%%%%%%%%%%%%%%%%%%%%%%%%%70

\section{Preliminaries}
\label{sec-preliminaries}
%\mbox{}

Let us consider the kaon-nucleon elastic scattering process:
\begin{equation}
N_\alpha(k,\lambda) \ \ + \ \ K_i(p) \ \
\rightarrow \ \
N_\beta(k',\lambda') \ \ + \ \ K_j(p'). \ \
\label{1a}
\end{equation}
Here
$\lambda, \; \lambda' = \{1,2\}$
stand for the nucleon spin variables and
$i, j \, (\alpha,\beta) = \{1,2\}$
are the kaon (nucleon) isotopic indices
(see Appendix~\ref{app1}
for the summary of our field conventions). The isotopic structure of
the amplitude  reads:
\be
% &&
M_{\alpha i}^{\ \ \ \beta j} = i(2 \pi)^4 \delta^{(4)}(p+k-p'-k')
 %\nonumber \\&&  \times
 \big\{ \delta_{\alpha\, \cdot}^{\cdot \, \beta} \delta_{i \, \cdot}^{\cdot \, j}
 M^{+} (\lambda, \lambda')+
 \delta_{\alpha \, \cdot}^{ \cdot \, j} \delta_{i \, \cdot}^{\cdot \, \beta}
 M^{-} (\lambda, \lambda') \big\}\,.
\ee
Each one of the isotopic amplitudes
$M^\pm$
can be presented as follows:
\be
%&&
 M^{\pm }(\lambda, \lambda',s,t,u) % \nonumber \\
%&&
=\overline{u}(\lambda',k')
 \left\{
 A^\pm(s,t,u)+\hat{Q}B^\pm(s,t,u)
 \right\}
 u(\lambda,k)\, .
 \label{invAml}
\ee
Throughout the paper we adopt the Dirac ``hat''  notation:
$\hat{p} \equiv p^\mu \gamma_\mu$;
$\overline{u}(k',\lambda')$, $u(k,\lambda)$
stand for the nucleon Dirac spinors and
$ Q \equiv \frac{p+p'}{2}$.
The invariant amplitudes
$A^{\pm}$ and $B^{\pm}$
may be considered depending on arbitrary pair of the Mandelstam
variables:
\be
&& s=(k+p)^2, \ \ t=(p-p')^2, \ \ u=(k-p')^2;  \nonumber
\\&& s+t+u=2(m^2+\mu^2)
\equiv 2 \sigma\, ,
\nonumber
\ee
where
$m(\mu)$
stands for the nucleon (kaon) mass. We introduce the  special
notations for two useful combinations of mass parameters:
$$
\theta \equiv (M^2_{\sss R} - \sigma),\ \ \ \ \ \
\Sigma \equiv (M^2_{\sss R} - 2 \sigma )\, ,
$$
where
$M_R$
is the mass parameter of a resonance
$R$.

\begin{figure}
\centering
\includegraphics[height=2.5cm]{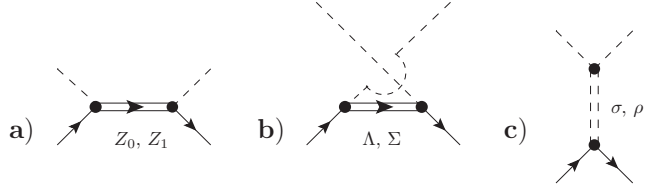}
\caption{We need to calculate the on-shell numerators
of these graphs.}
\label{2f}
\end{figure}

The construction of scattering amplitudes to a given loop order
in the EST approach implies the use of the modified system of Feynman
rules, containing only minimal (resultant) vertices and minimal
propagators. Our present goal is to work out the tree level
expressions for
$A^\pm$
and
$B^\pm$.
To make use of the technique of Cauchy forms for these amplitudes we
need to specify their residues at poles corresponding to
$s$-, $t$- and $u$-channel
resonance exchanges. In other words, we have to calculate the on-shell
numerators in the expressions which correspond to the contributions of
graphs shown on
Fig.~\ref{2f}.
For this we need to compute the products of the form
$V \Pi V$,
where
$V$
stand for the relevant minimal triple vertices while
$\Pi$
denotes the covariant spin sum (numerator of the minimal propagator)
of a resonance in question.

Analogous to the case of
$\pi N$
scattering
\cite{KSAVVV3},
it is possible to construct the corresponding minimal triple vertices
with the help of the listed below Hamiltonian monomials
(c.f. \cite{Car}).
Let us stress that this is only possible with respect to minimal
vertices with three lines. The parametrization of minimal vertices
with
$l \geq 4$
lines could be written out explicitly only in the momentum space.

Let us first consider the monomials that correspond to the depicted on
Fig.~\ref{2f}b
$u$-channel
baryon resonance
$R_u$
exchanges with strangeness
$S=-1$
and isospin
$I=0,1$
(in particular  $\Lambda(1115)$ and $\Sigma(1189)$ hyperons
occur in this channel). They look as follows:
\be
%&&
 H(R_u \bar{K} N)
% \nonumber \\ &&
= g_{\sss R} {P^{u \, (0,1) }}^{\cdot \, k \, \alpha \, \cdot}_{\rho \, \cdot \; \cdot \,  j} \,
 \overline{\Psi}_\alpha
 \Gamma(\mathcal{N})
 {R^{\rho\, \cdot}_{\cdot\,k}}_{   \mu_1...\mu_l}  \partial^{\mu_1...\mu_l}K^j+
       {\rm h.c.}  \, .
\nonumber \\ &&
\label{lagZ}
\ee
Here
${P^{u \, (0,1) }}^{\cdot \, k \, \alpha \, \cdot}_{\rho \, \cdot \; \cdot \,  j}$
are the
$u$-channel
isospin projectors
(see (\ref{prjU}))
while
\begin{equation}
\Gamma(\mathcal{N}) =
\begin{cases}
1_{4\times4}, & \text{for} \;  \mathcal{N}=-1 , \\ i\gamma_5,
 & \text{for} \; \mathcal{N}=+1 ;
 \end{cases}
 \label{mtrx}
\end{equation}
the notation
$\mathcal{N} \equiv \mathbf{P}(-1)^l$
is used for the normality of a resonance with spin
$j=l+\frac{1}{2}$
and parity
$\mathbf{P}$;
$g_{\sss R}$
stands for the minimal (dimensional) coupling constant.

The Hamiltonian monomials for the vertices describing
$s$-channel
exchanges (with strangeness
$S=1$
exotic baryon resonances) only differ from
(\ref{lagZ})
by the isospin projectors
(\ref{prjS}).

We also need the Hamiltonian monomials which correspond to minimal
vertices describing
$t$-channel
non-strange meson exchanges with isospin
$I=0,\,1$,
normality
$ \mathcal{N}=+1$
and spin
$j=l$.
There are two monomials of this kind since the minimal
$RN\bar{N}$
vertex contains two independent tensor structures:

\be
&&
H(R_t N \bar{N})=
{P^{t \, (0,1) }}^{\cdot \, \sigma \, n \, \cdot}_{\rho \, \cdot \; \cdot \,  i}
       \left \{
        \frac{1}{2} f^{(1)} g^{(1)}_{\sss RN\overline{N}}
         \overline{\Psi}_\sigma
          \partial^{\mu_1...\mu_l}
           \Psi^\rho \right. \nonumber \\ &&
          + \left.
            \frac{1}{2} f^{(2)} g^{(2)}_{\sss RN\overline{N}}j
             \overline{\Psi}_\sigma
              \gamma_{\mu_j}
               \partial^{\mu_1...\mu_{l-1}}
                \Psi^\rho
        \right\}
                {R^{ \cdot \, i}_{n\, \cdot  }}_{  \mu_1...\mu_l} + {\rm h.c.} \nonumber \\ &&
 \label{MesonLagr}
H(R_t K \bar{K})
\nonumber \\ &&
=i\, \frac{1}{2}\, g_{\sss RK\overline{K}}
    % P^{(0,1) \;  \sigma n}_{\ \ \ \rho \ \ \ \ \ i}
    {P^{t \, (0,1) }}^{\cdot \, \sigma \, n \, \cdot}_{\rho \, \cdot \; \cdot \,  i}
      {R^{\rho\, \cdot}_{\cdot \, \sigma}}_{  \; \mu_1...\mu_l}\overline{K}^i
       \partial^{\mu_1...\mu_l}K_n +  {\rm h.c} \, . \nonumber \\ &&
\ee
Here
${P^{t \, (0,1) }}^{\cdot \, \sigma \, n \, \cdot}_{\rho \, \cdot \; \cdot \,  i}$
stand for $t$-channel isospin projectors
(\ref{prjT}).
Besides, we introduce the phase factors
$f^{(1,2)}$
\begin{equation}
f^{(1)}=
\begin{cases}
1, & \! \! \text{for} \;  J=0,2,..  \\ i,
 & \! \! \text{for} \; J=1,3,...
 \end{cases};
 \ \
 f^{(2)}=
\begin{cases}
i, & \! \! \text{for} \;  J=0,2,.. \\ 1,
 & \! \! \text{for} \; J=1,3,...
 \end{cases}.
 \label{factors}
 \nonumber
\end{equation}
in order to ensure hermiticty.

The numerator of minimal propagator is just
a covariant spin sum.
So, minimal propagators for
$s$- and $u$-channel
baryon resonances read:
\be
 \frac{i}{(2 \pi)^4}\,
\frac{
%(-1)^l
\Pi_{\xi \, \mu_1...\mu_l; \; \eta \, \nu_1...\nu_l }(q)}
{q^2-M_R^2+ i \epsilon}, \label{spin-sum-bary}
\ee
where
$\Pi_{\xi\, \mu_1...\mu_l; \; \eta \, \nu_1...\nu_l}(q)$
is the covariant spin sum of the spin
$j=l+ \frac{1}{2}$
baryon resonance
\cite{Alfaro}
($\xi, \, \eta= \{1,2,3,4\}$
stand for the Dirac spinor indices).
The minimal propagator for
$t$-channel
meson resonance differs from
(\ref{spin-sum-bary})
only by the structure of spin sum.

The technique needed to calculate the principal parts of contributions
of the graphs with
spin-$j$
resonance exchanges is described in
\cite{KSAVVV3};
here we briefly illustrate it by way of computing the graph shown in
Fig.~\ref{2f}~(b).
Using the minimal vertices that correspond to the monomial
(\ref{lagZ}),
we obtain the following expression for the matrix element of
$u$-channel
exchange by the baryon resonance
$R$
with the mass parameter
$M_R$,
isospin
$I$
and normality
$ \mathcal{N} $:
\begin{equation}
-
%i(2 \pi)^4 \delta^{(4)}(p+k-p'-k')
{P^{u \, (0,1) }}^{\cdot \, j \, \beta \, \cdot}_{\alpha \, \cdot \; \cdot \,  i} \,
 %P_{\ \ \alpha i}^{(I) \;  \beta j }
 g_R ^2
  \overline{u}^+(\lambda',k') \Gamma
   \frac{(-1)^l \mathcal{P}_{l+\frac{1}{2}} (-p,-p')}{u-M_R^2}
    \Gamma u^-(\lambda,k).
\label{s-channel-GR}
\end{equation}
Here
${P^{u \, (0,1) }}^{\cdot \, j \, \beta \, \cdot}_{\alpha \, \cdot \; \cdot \,  i}$
stands for the
$u$-channel
isospin projector
(\ref{prjU})
and
$\mathcal{P}_{l+\frac{1}{2}} (-p,-p')$ is
the contracted projector
(covariant spin sum contracted with the appropriate number of
$-p$
and
$-p'$
vectors).

Since all we need are the expression for the residue at
$u=M^2_{\sss R}$
we make use of an explicit form
\cite{Alfaro}
for the contracted projector
$\mathcal{P}_{l+\frac{1}{2}} (p',p)$
calculated between the nucleon spinors under the on-mass-shell (OMS)
conditions
$(k^2=k'^2=m^2, \, p^2=p'^2=\mu^2, \, u=M_{\sss R}^2)$:
\be
&&
\overline{u}^+(\lambda',k') \mathcal{P}_{l+\frac{1}{2}}
(-p,-p')%&
\,  u^-(\lambda,k) \Big|_{\sss OMS}  \nonumber \\ &&
 =\overline{u}^+(\lambda',k')
 %\nonumber \\ &&
\frac{l! (-1)^l}{(2l+1)!!} \,
\left[ F^l_A(- \mathcal{N}M_R,t)+
\hat{Q} F^l_B(- \mathcal{N}M_R,t)
\right]
u^-(\lambda,k).
%\nonumber \\ &&
\label{Conpr}
\ee
Here
$F^l_{A,B}$
are given by the relations
(\ref{Fa}, \ref{Fb}) (see Appendix
\ref{app2}).
With the help of
(\ref{Conpr})
one can easily calculate the contributions of
(\ref{s-channel-GR})
to the principle parts of invariant amplitudes
$A^\pm$
and
$B^\pm$
defined in
(\ref{invAml}).

Computing all the elements shown on
Fig.~\ref{2f}
one can gather the contributions of individual graphs into four
invariant amplitudes
$ X^\pm=\{A^\pm, \; B^\pm \}$
and write down the final result for their principal parts. In terms of
shortened notations introduced in
Appendix \ref{app2}
the formal expressions for the principal parts of the invariant
amplitudes read:
\be
{\bf P.p.} [ \, X^\pm(s,t,u)] %&& \nonumber \\
&&
=-\sum_{{B(I=0,1)} \atop {S=+1}}c_I^\pm
\frac{Y_X(...,t)|_{s=M_s^2}}{s-M_s^2}  % \nonumber \\ &&
-\sum_{{B(I=0,1)} \atop {S=-1}} \eta_{\sss X} b_I^\pm
\frac{Y_X(...,t)|_{u=M_u^2}}{u-M_u^2}  \nonumber \\ &&
-\sum_{{M(I=0,1)} \atop {S=0}}d_I^\pm
\frac{W_X(...,\frac{s-u}{4F})|_{t=M_t^2}}{t-M_t^2}\, .
\label{formalAB}
\ee
The first two sums are taken over all possible baryon resonances with
isospin
$I=0,1$
and strangeness
$S= \pm 1$.
The third sum is taken over all possible non-strange meson resonances
with isospin
$I=0,1$.
Let us stress that at this stage the  sums in
(\ref{formalAB})
are to be taken just as
formal.
The  formulation of suitable convergency conditions for these formal
series allows one to define rigorously the
$0$th
(tree level) order approximation of the loop expansion for the EST amplitude in
the sector of strange hadrons.

\section{Cauchy forms in three hyperlayers}
\label{sec-Cforms}

\begin{figure}
\centering
\includegraphics[height=6.0cm]{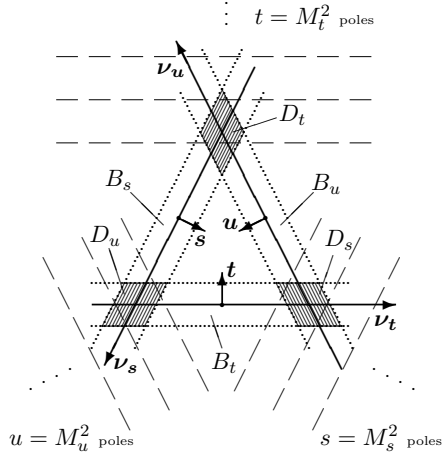}
\caption{   Mandelstam plane: three different Cauchy series converge
uniformly in three different hyperlayers
$B_s$, $B_t$
and
$B_u$
(their sections by the Mandelstam plane are bounded by dotted lines);
$(s,\; \nu_s\equiv u-t)$,
$(t,\; \nu_t\equiv s-u)$,
$(u,\; \nu_u\equiv t-s)$
are the natural coordinates in
$B_s$,
$B_t$
and
$B_u$,
respectively. The domains of the layer intersections are hatched
(denoted as
$D_s$,
$D_t$,
and
$D_u$).
Approximate positions of the poles (pole lines) in
$s$,
$t$,
and
$u$-channels
are shown by dashed lines (the mass parameters are real).
\label{5f} }
%\end{center}
\end{figure}

In accordance with the summability principle
\cite{KSAVVV2},
within our EST approach the tree level invariant amplitudes of
$KN$
scattering
$X^\pm= \{ A^\pm, \, B^\pm \}$
are required to be meromorphic functions in each pair energy
$\nu_x$
($\nu_s\equiv u-t$,
$\nu_t\equiv s-u$,
$\nu_u\equiv t-s$)
at arbitrary fixed value of the  momentum transfer
$x=\{s,\,t,\,u\}$.
The uniformity principle specifies that in every hyperlayer
$B_x: \;  \{\nu_x \in \mathbb{C}, \;
%x \in \mathbb{R}, \;
x \sim 0 \}$
(see Fig.~\ref{5f})
containing the zero momentum transfer ($x=0$) hyperplane, the
invariant amplitudes must be polynomially bounded functions of the
corresponding variable
${\nu}_x$.
 The bounding polynomial degree in every hyperlayer
$B_x$
is fixed by the value of the relevant Regge intercept.
The method of the Cauchy forms (that is in fact the adaptation of the
conventional dispersion relation technique for the case of meromorphic
functions) allows one to present the amplitude
which is
$N$-bounded
in a hyperlayer
$B_x$
as a uniformly converging series of pole contributions.

%%%%%%%%%%%%%%%%%%%%%%%%%%%%%%%%%%%%%%%%%%%%%%%%%%%%%%%%%%%%%%%%%%%%%%
%%%%%%%%%%%%%%%%%%%%%%%%%%%%%%%%%%%%%%%%%%%%%%%%%%%%%%%%%%%%%%%%%%%%%%

Let us start with constructing the Cauchy forms in the hyperlayers
$B_s$
and
$B_u$.
The situation in these cases is trivial since in both layers all the
invariant amplitudes possess decreasing asymptotic behavior
(see Appendix~\ref{appA3}).
Therefore the relevant Cauchy forms are just  sums of pole
contributions. This information together with the formal expressions
(\ref{formalAB})
for principle parts of the invariant amplitudes is sufficient for
constructing the well defined Cauchy forms for tree level invariant
amplitudes of
$KN$
scattering in these layers. The only singularities of the
\emph{tree level}
graphs in these hyperlayers are simple poles in variables
$\nu_s$
and
$\nu_u$.

With the use of the compact notations introduced in
Appendix~\ref{app2}
one may treat all
$X^\pm$
on the same footing.
The residues at poles which correspond to the
$u$-channel
baryon resonances with strangeness
$S=-1$
and hypothetic exotic
$s$-channel
resonances with
$S=+1$
are given by
$Y_X$
(\ref{Y_X}).
The residues  at poles corresponding to the
$t$-channel
meson resonance exchanges are given by the functions
$W_{\sss X}$
(\ref{WaWb}).

Thus in the layer
$B_s$
we have:
\be
 X^\pm (s, \, \nu_s)     &&
 =-\sum_{{B(I=0,1)} \atop {S=-1}} \eta_X b_I^\pm
 Y_X\left( ...,-(\Sigma+s)\right)
 \frac{2}{\nu_s-(s+2\theta)}
 \nonumber \\ &&
 -\sum_{M(I=0,1)} d_I^\pm W_X\left( ...,\frac{2s+\Sigma}{4F}\right)
 \frac{-2}{\nu_s+(s+2\theta)}\, ,
% \nonumber \\ &&
 \label{CformBs}
\ee
while in
$B_u$:
\be
X^\pm (u, \, \nu_u)  &&
=-\sum_{{B(I=0,1)} \atop {S=+1}} \eta_X b_I^\pm
Y_X\left( ...,-(\Sigma+u)\right)
\frac{-2}{\nu_u+(u+2\theta)}
     \nonumber \\ &&
- \sum_{M(I=0,1)} d_I^\pm W_X\left( ...,\frac{-(2u+\Sigma)}{4F}\right)
\frac{2}{\nu_u-(u+2\theta)}\, .
\nonumber \\ &&
\label{CformBu}
\ee
In what follows we also employ the Cauchy forms
(\ref{CformBs})
and
(\ref{CformBu})
rewritten in terms of corresponding natural variables in the domains
$D_u = B_s\cap B_t$,
$D_t = B_u\cap B_s$,
$D_s = B_t\cap B_u$.

Let us now consider the hyperlayer $B_t$.
%$B_t: \, \{ t \sim 0, \; \nu_t=s-u \in \mathbb{C} \}.$
The Regge theory requirements listed in the
Appendix~\ref{appA3}
provide %the following
information on the asymptotic behavior of
the isotopic combinations of the invariant amplitudes
$(2X^+ + X^-)$
and
$X^-$
for large
$|\nu_t|$.
%
%:
%\begin{itemize}
%\item
%$(2A^+ +A^-)$
%is
%$1$-bounded;
%\item
%$A^-$
%and
%$(2B^++B^-)$
%are
%$0$-bounded;
%\item
%$B^-$
%has decreasing asymptotics.
%\end{itemize}
Thus in the layer
$B_t$
the Cauchy forms for tree level amplitudes
%$X^\pm$ $(X=A,B)$
require introducing the correcting polynomials in
$\nu_t$
of the degrees
$N_t$:
\be
&& N_t(2A^++A^-) = 1\, ;\ \ \ \ \ \
N_t(A^-) = 0 \,;
\nonumber \\ &&
N_t(2B^++B^-) = 0 \,;\ \ \ \ \ \
N_t(B^-) = -1 \, .
\label{t-intercepts}
\ee
Surely, the smooth terms -- polynomials of the same degrees in
$\nu_t$
with the coefficient functions depending on
$t$ --
must be taken into account.

For
$A^+$
in
$B_t$
we put down the Cauchy form with the
background term and correcting polynomials   of $1$st
order in $\nu_t$:
\begin{equation}
A^+(t, \nu_t)= \alpha_{\sss A^+}^{0}(t)
+\alpha_{\sss A^+}^{1}(t)\nu_t + \widetilde{A}^+(t,\nu_t).
\label{CformA+Bt}
\end{equation}
Here by
$\widetilde{A}^+$
we denote the principle part of the corresponding Cauchy form with the
necessary correcting polynomials:
\be
\widetilde{A}^+(t,\nu_t)     &&
\equiv  - \sum_{{B(I=0,1)} \atop {S=+1}} Y_A(...,t)
\left(
   \frac{2c_I^+}{\nu_t-(t+2\theta)}+\frac{2c_I^+}{t+2 \theta}+
   \frac{2 \nu_t(c_I^+ + \frac{1}{2}c_I^-)}{(t+2 \theta)^2}
\right)  \nonumber \\
&& - \sum_{{B(I=0,1)} \atop {S=-1}} Y_A(...,t)
\left(
   \frac{-2b_I^+}{\nu_t+(t+2\theta)}+\frac{2b_I^+}{t+2 \theta}-
   \frac{2 \nu_t(b_I^+ + \frac{1}{2}b_I^-)}{(t+2 \theta)^2}
\right)
\nonumber \\ &&
\label{A+Bt-sing-part}
\ee
and
$\alpha_{\sss A^+}^{0,1}(t)$
describe the regular at
$t=0, \, \nu_t=0$
part of
$A^+$.

The Cauchy forms for
$A^-$
in
$B_t$
must contain the correcting polynomials and background terms of
$0$th
degree in
$\nu_t$,
therefore:
\be
&&
A^-(t, \nu_t)= \alpha_{\sss A^-}^{0}(t) + \widetilde{A}^-(t,\nu_t)\, .
\label{CformA-Bt}
\ee
Here, as above, the notation
\be
   \widetilde{A}^- (t,\nu_t)
&&
\equiv   -\sum_{B(I=0,1)\atop{S=+1}}  c_I^- Y_{A} (...,t)
\left(
\frac{2}{\nu_t-(t+2\theta)}+\frac{2}{t+2 \theta}
\right)
\nonumber \\&&
- \sum_{B(I=0,1)\atop{S=-1}}   b_I^-  Y_{A}(...,t)
\left(
\frac{-2}{\nu_t+(t+2\theta)}+\frac{2}{t+2 \theta}
\right).
%\nonumber \\ &&
\label{A-Bt-sing-part}
\ee
is used for the singular (``principal'') part of the Cauchy form with
necessary correcting polynomials.

The Cauchy form for
$B^+(t,\nu_t)$
is similar to that for
$A^-(t,\nu_t)$
(it also requires introducing the correcting polynomials of
$0$th
degree in
$\nu_t$):
\begin{equation}
B^+(t, \nu_t)= \alpha_{\sss B^+}^{0}(t) + \widetilde{B}^+(t,\nu_t)\, .
\label{CformB+Bt}
\end{equation}
Here
\be
\widetilde{B}^+(t,\nu_t)  &&
\equiv   -\sum_{B(I=0,1)\atop{S=+1}}   Y_B(...,t)
\left(
\frac{2 c_I^+}{\nu_t-(t+2\theta)}+\frac{2c_I^+ + c_I^-}{t+2 \theta}
\right)
\nonumber \\ &&
+ \sum_{B(I=0,1)\atop{S=-1}}    Y_B(...,t)
\left(
\frac{-2 b_I^+}{\nu_t+(t+2\theta)}+\frac{2b_I^++b_I^-}{t+2 \theta}
\right).
%\nonumber \\ &&
\label{B+Bt-sing-part}
\ee

Finally, the Cauchy form for
$B^-(t,\nu_t)$
in
$B_t$
is just the sum of pole contributions:
\be
 B^-(t, \nu_t)  &&
 = -\sum_{B(I=0,1)\atop{S=+1}}   Y_B(...,t)
 \frac{2c_I^-}{\nu_t-(t+2\theta)}
 \nonumber \\ &&
 -
 \sum_{B(I=0,1)\atop{S=-1}}   Y_B(...,t)
 \frac{2b_I^-}{\nu_t+(t+2\theta)} \equiv \widetilde{B}^-(t,\nu_t) .
 \label{CformB-Bt}
\ee

To derive the system of bootstrap constrains we need to rewrite the
Cauchy forms for
$X^\pm$
in terms of natural variables of relevant intersection domains.

For example, in the intersection domain
$D_u \equiv B_t \cap B_s$
the natural variables are
$t$
and
$s$
(both
$t,s \sim 0$).
Therefore making use of the relation
$
\nu_t \equiv s-u = t + 2s - 2\sigma
$
we obtain
\be
&&
A^+ \big|_{D_u} \! (t,s)
= \alpha_{\sss A^+}^{0}(t)
+ \alpha_{\sss A^+}^{1}(t)(t+2s-2 \sigma) + \widetilde{A}^+(t,s)\, ;
%\label{hat}
\nonumber \\ &&
A^-\big|_{D_u} \! (t,s)= \alpha_{\sss A^-}^{0}(t)
 +\widetilde{A}^-(t,s);   \nonumber \\ &&
B^+\big|_{D_u} \! (t,s)=
\alpha_{\sss B^+}^{0}(t) +\widetilde{B}^+(t,s);  \nonumber \\ &&
 B^-\big|_{D_u} \! (t,s) = \widetilde{B}^-(t,s)\, .
 \label{hatA+A-B+B-}
\ee
Here
$ \widetilde{X}^\pm$ $(X=A,B)$
stand  for the singular (``principal'') parts of corresponding Cauchy
forms with necessary correcting polynomials rewritten in terms of
natural  variables of the given intersection domain. We adopt the
following convention on the order of arguments of
$\widetilde{X}^\pm$:
the natural variable which marks the hyperlayer where the initial
Cauchy form was written  (in the case under consideration this is
$B_t$)
stands at the first position.

%%%%%%%%%%%%%%%%%%%%%%%%%%%%%%%%%%%%%%%%%%%%%%%%%%%%%%%%%%%%%%%%%%%%70

%%%%%%%%%%%%%%%%%%%%%%%%%%%%%%%%%%%%%%%%%%%%%%%%%%%%%%%%%%%%%%%%%%%%%%

\section{The structure of bootstrap equations}
\label{sec-strboot}
%\mbox{}

The tree level bootstrap conditions follow from the requirement that
every tree level invariant amplitude must be a meromorphic function
with definite asymptotic behavior in hyperlayers $B_s$, $B_t$
and
$B_u$. Hence
three different Cauchy forms which present the amplitude in these layers
must coincide pairwise in the relevant intersection domains
$D_s$, $D_t$
or
$D_u$.
As pointed out in
\cite{KSAVVV2},
the bootstrap conditions restrict the allowed values of the tree-level
resultant parameters. It is these parameters which stand in the right
hand sides of renormalization prescriptions fixing the physical
content of effective scattering theory in the renormalization scheme
without oversubtractions. %Because of this reason they present physical
%(measurable) quantities.
 Once resolved the (full) system of tree level
bootstrap constraints would single out the set of essential parameters
of a theory. The higher level bootstrap conditions only can further
restrict this set.

Let us first construct the system of tree level bootstrap conditions
for the invariant amplitude
$A^+$.
In each one of three intersection domains
$D_u = B_s \cap B_t$,
$D_t = B_u \cap B_s$
and
$D_s = B_t \cap B_u$
the two different Cauchy series for
$A^+$
(see
Section~\ref{sec-Cforms})
are equally applicable. Thus employing the conventions of
Section~\ref{sec-Cforms}
and the notations introduced in
\Ref{hatA+A-B+B-} we have:
\begin{itemize}
\item In $D_u$ (for $s,t \sim 0$):
\be
&&
\begin{cases}
A^+\big|_{\sss D_u} \! \!= {A}^+(s,t)\,; %\big| _{\sss B_s}
\\
A^+\big|_{\sss D_u} \! \!= \alpha_{\sss A^+}^{0}(t)+
\alpha_{\sss A^+}^{1}(t)(t+2s-2 \sigma)+
\widetilde{A}^+(t,s)\, . %\big|_{\sss B_t}
\end{cases}
\nonumber %\\ &&
\ee
\item In $D_t$ (for $s,u \sim 0$):
\be
%D_t\{s,u \sim 0\}:\ \ \ \
%\left\{
\begin{cases}
 A^+\big|_{\sss D_t} = {A}^+(s,u)\,; %\big|_{B_s}
  \\
  A^+\big|_{\sss D_t} = {A}^+(u,s). %\big|_{B_u}
\end{cases}
\nonumber
\ee
\item In $D_s$ (for $t,u \sim 0$):
\be
&&
\begin{cases}
 A^+\big|_{\sss D_s}  \! \! =  {A}^+(u,t) \,; %\big|_{B_u}
 \\
 A^+\big|_{\sss D_s} \! \! = \alpha_{\sss A^+}^{0}(t)-
\alpha_{\sss A^+}^{1}(t)(t+2u-2 \sigma)+
\widetilde{A}^+(t,u). %\big|_{B_t}
\end{cases}
\nonumber %\\ &&
\ee
\end{itemize}

To ensure the possibility of analytic continuation from one hyperlayer
to another, each pair of the relevant series must coincide identically
in the intersection domain where both expansions are valid. Thus we
obtain the following system of conditions:
\begin{itemize}
\item In $D_u$:
\be
 \alpha_{\sss A^+}^{0}(t)+
\alpha_{\sss A^+}^{1}(t)(t+2s-2\sigma) +\varphi_{\sss A^+} (s,t)=0,
\label{17.a}
\ee
where $\varphi_{\sss A^+} (s,t) \equiv \widetilde{A}^+(t,s) -
 {A}^+(s,t)$.
\item In $D_t$:
\be
 0= \Phi_{A^+} (u,s) \equiv   {A}^+(s,u)-  {A}^+(u,s)\,;
\label{17.b}
\ee

\item In $D_s$:
\be
\alpha_{\sss A^+}^{0}(t)+
\alpha_{\sss A^+}^{1}(t)(-t-2u+2\sigma)-\Psi_{\sss A^+} (t,u)=0,
%\nonumber \\ &&
\label{17.c}
\ee
where $\Psi_{\sss A^+} (t,u) \equiv  {A}^+(u,t) - \widetilde{A}^+(t,u)$.
\end{itemize}
Here we have introduced three
{\em generating functions}:
$\varphi_{\sss A^+}(s,t)$,
$\Phi_{\sss A^+}(u,s)$
and
$\Psi _{\sss A^+}(t,u)$
(see Appendix~\ref{app3})
which are the differences of sums of principle parts (with correcting
polynomials, if needed) of two Cauchy forms in the corresponding
hyperlayers. In the same way as in
\cite{AVVV1}, \cite{KSAVVV2}
we exploit the fact that in
(\ref{17.a})--(\ref{17.c})
the dependence of generating functions (or their partial derivatives)
on certain Mandelstam variables is purely fictitious. This allows one
to express explicitly the unknown functions
$\alpha$
in terms of spectrum parameters (in
\cite{KSAVVV2}
such expressions were called
{\em  the first kind bootstrap constraints}).
Besides, we derive the consistency conditions for these expressions
({\em second kind bootstrap constraints}).
Throughout the text we adopt the following notations for partial
derivatives in the Mandelstam variables
$x=\{s,\,t,\,u\}$:
\be
(\partial_x)^k \equiv \frac{\partial^k}{\partial x^k}\,.
\ee

First we note that, according to
\Ref{17.b},
$\Phi_{\sss A^+}(u,s)$
is identically zero everywhere in the vicinity of the point
$u=0, \;s=0$.
Therefore the following consistency conditions hold:
\begin{equation}
(\partial_u)^p (\partial_s)^k
\Phi_{\sss A^+}(u,s)\Big|_{{u=0} \atop {s=0}}=0,
  \ \   { \rm for \; all}  \ \ p,k=0,1,2,...   \ \ .
\label{consPhiA+}
\end{equation}

Next, from
(\ref{17.c})
for all
$t,u \sim 0$
we define the unknown background term:
\begin{equation}
\alpha_{\sss A^+}^{1}(t)=-\, \frac{1}{2}\,
(\partial_u)\,
\Psi_{\sss A^+} (t,u)\, .
\label{alpha1+(t)}
\end{equation}
The condition which ensures the consistency of the
definition
(\ref{alpha1+(t)})
reads:
\begin{equation}
    (\partial_t)^p (\partial_u)^{k+2}
   \Psi_{\sss A^+} (t,u)\Big|_{{t=0} \atop {u=0}}=0,  \ \     { \rm for \; all}  \ \  p,k=0,1,2,...   \ \ .
 \label{consPsiA+}
\end{equation}
At the same time, the same function
$\alpha_{\sss A^+}^{1}(t)$
can be defined from
(\ref{17.a}):
\begin{equation}
\alpha_{\sss A^+}^{1}(t)=-\, \frac{1}{2}\,
%\frac{\partial}{\partial s}\,
%(\partial/\partial s) \,
(\partial_s)
\varphi_{\sss A^+}(s,t)\, .
\label{alpha1+(t)alternative}
\end{equation}
The consistency condition for this -- alternative -- definition reads:
\begin{equation}
%\frac{\partial^{p+k+2}}{\partial s^{p+2} \partial t^k}
(\partial_s)^{p+2} (\partial_t)^k
\varphi_{\sss A^+}(s,t)\Big|_{{s=0} \atop {t=0}}=0,  \ \   { \rm for \; all}  \ \  p,k=0,1,2,...  \ \   .
\label{consvarphiA+}
\end{equation}
From the requirement that the expression
(\ref{alpha1+(t)})
should not contradict to another -- equally possible -- expression
\Ref{alpha1+(t)alternative}
for the same function, we obtain the following system of equivalence
conditions:
\be
&&
%\frac{\partial^p}{\partial t^p}
\! \! \!
(\partial_t)^p
\Big[
%\frac{\partial}{\partial u}
(\partial_u)
\Psi_{\sss A^+}(t,u)_{u=0}
\Big]_{t=0} \! \!=
%\frac{\partial^p}{\partial t^p}
(\partial_t)^p
\Big[
%\frac{\partial}{\partial s}
(\partial_s)
\varphi_{\sss A^+}(s,t)_{s=0}
\Big]_{t=0}\,.
\nonumber \\ &&
{ \rm for \; all}  \ \  p=0,1,2,... \,:
\label{concalpha1+}
\ee

Finally, from
\Ref{17.c}
and
\Ref{alpha1+(t)}
we define for all
$t \sim 0$
the second unknown function
$\alpha_{\sss A^+}^{0}(t)$:
\begin{equation}
\alpha_{\sss A^+}^{0}(t)=
\Psi_{\sss A^+}(t,0)-\frac{1}{2}\, (t-2\sigma)
\Big[
%\frac{\partial}{\partial u}
(\partial_u)
\, \Psi_{\sss A^+}(t,u)\Big]_{u=0}\, .
\label{alpha0+(t)1}
\end{equation}
Alternatively, this function can be derived from
\Ref{17.a}:
\begin{equation}
\alpha_{\sss A^+}^{0}(t) = \varphi_{\sss A^+}(0,t) +
\frac{1}{2}\, (t-2\sigma)
\Big[
%\frac{\partial}{\partial s}\,
(\partial_s)
\varphi_{\sss A^+}(s,t)
\Big]_{s=0}\, .
\label{alpha0+(t)2}
\end{equation}
The corresponding system of equivalence conditions reads:
\begin{multline}
%\frac{\partial^p}{\partial t^p}
(\partial_t)^p
\Big\{
\frac{(t-2\sigma)}{2}\,
\Big[
%\left.
%\frac{\partial}{\partial s}
(\partial_s)
\varphi_{\sss A^+}(s,t) \Big|_{s=0} +
%\frac{\partial}{\partial u}
(\partial_u)
\Psi_{\sss A^+}(t,u) \Big|_{u=0}
\Big]
\\
+ \Big[ \varphi_{\sss A^+}(0,t) - \Psi_{\sss A^+}(t,0) \Big]
\Big\}
=0, \ \   { \rm for \; all}  \ \  p=0,1,2,..  \ \ .
\label{concalpha0+}
\end{multline}

Bootstrap constraints for
$A^-$
may be derived in the same way; they are more simple because
$A^-$
possesses constant asymptotics in the hyperlayer
$B_t$.
As above, we introduce three generating functions (see
Appendix~\ref{app3}):
$\varphi_{\sss A^-}(s,t)$,
$\Phi_{\sss A^-}(u,s)$,
and
$\Psi_{\sss A^-}(t,u)$.
Then, the set of bootstrap constraints contains one condition of the
first kind
\begin{equation}
\alpha_{\sss A^-}(t)=\Psi_{\sss A^-}(t,0)
\label{alpha0-(t)}
\end{equation}
(it fixes the unknown function
$\alpha_{A^-}(t)$)
and four systems of the second kind constraints. Three of these latter
systems, namely,
\begin{equation}
\begin{split}
&
%\frac{\partial^{p+k+1}}{\partial t^p \partial u^{k+1}}
(\partial_t)^p (\partial_u)^{k+1}
\Psi_{\sss A^-}(t,u)\Big|_{{t=0} \atop {u=0}}=0\, ,
  \\ &
(\partial_s)^{p+1} (\partial_t)^k
\varphi_{\sss A^-}(s,t)\Big|_{{s=0} \atop {t=0}}=0\, ,    \\ &
(\partial_u)^p (\partial_s)^k
\Phi_{\sss A^-}(u,s)\Big|_{{u=0} \atop {s=0}}=0\, ,
\label{consA-}
\end{split}
\end{equation}
$(p,k=0,1,2,...)$
present the consistency conditions while the fourth one
\begin{equation}
(\partial_t)^p
\Big[ \Psi_{\sss A^-}(t,0)+
\varphi_{\sss A^-}(0,t)
\Big]_{t=0}=0\,; \ \   { \rm for \; all}  \ \ p=0,1,2,.. \;.
\label{concalpha0-}
\end{equation}
ensures the equivalence of two possible definitions of
$\alpha_{\sss A^-}(t)$
(it plays the same role as
%\Ref{consvarphiA+}
\Ref{concalpha1+}
and
\Ref{concalpha0+}).

To write down the bootstrap constraints for
$B^\pm$
we introduce six generating functions:
$\varphi_{\sss B^\pm}(s,t)$,
$\Phi_{\sss B^\pm}(u,s)$
and
$ \Psi_{\sss B^\pm}(t,u)$
(see
Appendix~\ref{app3}).
The constraints for
$B^+$
are analogous to those for
$A^-$.
The system for
$B^-$
is even more simple. It consists of three second kind bootstrap
conditions:
\begin{equation}
\begin{split}
&
 (\partial_t)^p (\partial_u)^k
 \Psi_{\sss B^-}(t,u)\Big|_{{t=0} \atop {u=0}}=0\,; \\ &
   (\partial_s)^p (\partial_t)^k
\varphi_{\sss B^-}(s,t)\Big|_{{s=0} \atop {t=0}}=0\,;  \\ &
  (\partial_u)^p (\partial_s)^k
  \Phi_{\sss B^-}(u,s)\Big|_{{u=0} \atop {s=0}}=0\,;
\ \   { \rm for \; all}  \ \ p=0,1,2,.. \;.
\end{split}
\label{consB-}
\end{equation}

Thus we have constructed the system of bootstrap conditions for the
invariant amplitudes of
$KN$
elastic scattering process. The first kind bootstrap constraints
\Ref{alpha1+(t)}, \Ref{alpha0+(t)1}
and
\Ref{alpha0-(t)}
define the smooth parts of those amplitudes. The constraints of the
second kind (namely,
\Ref{consPhiA+}, \Ref{consPsiA+},
\Ref{consvarphiA+}, \Ref{consA-}
and
\Ref{consB-})
provide the consistency conditions for these definitions. Finally,
(\ref{concalpha1+}), (\ref{concalpha0+})
and
(\ref{concalpha0-})
ensure that the definitions are not contradictive.

It should be noted that the above-obtained system of bootstrap
constraints still is not complete. The degrees of bounding polynomials
needed to construct the Cauchy forms for certain combinations of
invariant amplitudes in the layers
$B_s$
and
$B_u$
are
$\le -2$.
This results in appearing of additional super-convergence  conditions
(see
\cite{POMI})
which we are not going to consider here.

%%%%%%%%%%%%%%%%%%%%%%%%%%%%%%%%%%%%%%%%%%%%%%%%%%%%%%%%%%%%%%%%%%%%70

\section{Sum rules for $KN$ spectrum parameters }
\label{S_NumTests}
%\mbox{}

\subsection{Numerical testing of bootstrap constrains}
\label{Sec-5.1}

Tree level bootstrap equations derived in the previous Section
represent the set of limitations imposed by the requirement of
mathematical correctness of extended perturbation scheme
on the values of renormalization prescriptions (RPs) fixing the
physical content of EST. Therefore starting from the tree level
the bootstrap system results in non-trivial constraints for the
values of physical parameters of the theory. Higher level bootstrap
constraints obviously  differ from those of tree level; they may
impose additional constraints on the set of physical parameters.

In other words, within our EST approach the tree level bootstrap
constraints are valid at any order of loop expansion  and possess an
important predictive power. The numerical testing of tree level
bootstrap constraints is, therefore, rightful. Moreover, it is of
great interest because it allows one to make at least preliminary
conclusions about the consistency of basic postulates employed in
EST approach such as the summability and uniformity principles
with the present day phenomenology. Such a verification proved to be
successful in the cases of
$\pi K$
and
$\pi N$
scattering (see
\cite{KSAVVV3}, \cite{AVVV1}, \cite{AVPiN}).
Below we perform similar analysis of the bootstrap constraints for
the parameters of kaon-nucleon resonance spectrum.

Despite the fact that these parameters are known with much less
precision than those of pion-nucleon resonances, it still turns out
possible to single out the set of sum rules that are well saturated
by known experimental data. On the other hand, those sum rules which
are not so well saturated with now existing data, permit us to
speculate about possible scenarios that could amend the situation.
Thus in our numerical tests we also aim to show that the extended
perturbation scheme provides us with a tool to study the resonance
spectrum.

In our numerical studies we make use of the data
\cite{PDG}
on hadron spectrum (Table~1 of
Appendix~\ref{app4}).
Several phenomenological constants were taken from early reviews
\cite{Nagels,Dumbrajs:1983jd,EricsonWeise}.
The formulae connecting resonance couplings with the decay widths are
presented in
Appendix~\ref{app4}).

Note that we did not include in Table 1 the results of recent analysis
concerning the fine structure of the strange resonance spectrum (see,
e.g.
\cite{Oller}
and references therein). The reason is that the error bars induced by
the data on well established lowest resonances
$\Lambda$ and $\Sigma$
turn out to be larger than the possible total contribution of narrow
resonances discussed in
\cite{Oller}.
For this reason the latter contribution turns out invisible against
that background.

Obviously, the existing information on the
$KN$
resonance spectrum
\cite{PDG}
is incomplete in the region of high mass and spin. Moreover, much is
unclear with
$M>1 \, {\rm GeV}$
meson resonances in
$t$-channel
of the elastic
$KN$
reaction. Second,
spin-$\frac{1}{2}$
resonances over the
$u$-channel
threshold are not so well established too. One also needs to keep in
mind the possible existence of
$s$-channel
exotic resonances with strangeness
$S=+1$.
Therefore, our first goal is to find those sum rules which can be
saturated with the reliable experimental data. The invariant
amplitudes
$X^-=\{ A^-,\, B^-\}$
receive contributions from the exchanges with uncharged hyperons
in the
$u$-channel.
The main advantage is that both the
$\Lambda$ ($I=0$)
and
$\Sigma$ ($I=1$)
hyperon families contribute. Thus one may expect that due to mutual
cancelations the saturation of sum rules for
$X^-$
can be achieved faster than in the case of invariant amplitudes
$X^+$.
For the latter (as a consequence of isospin invariance) only the
$\Sigma$ ($I=1$)
family of hyperons contribute in the
$u$-channel.

As an example we have chosen the sum rules which follow from the
bootstrap constraints of the second kind
\Ref{consA-}
for the invariant amplitude
$A^-$.
\begin{equation}
(\partial_u)^p (\partial_s)^k
\Phi_{\sss A^-} (u,s)
\Big|_{u=0 \atop s=0}=0, \ \ \text{for all} \ \
p,\,k=0,1,... \;.
\label{SR_KN_Aminus_US}
\end{equation}
It turns out that for certain sum rules of this group the
contributions from some poorly established resonances are not
essential.

It is straightforward to check that for
$p=0,1,2$
and
$k=1,2$
the corresponding sum rules can be considered as purely baryonic ones.
Indeed in the meson sector only
isospin-$1$
resonances of odd spin
$J \ge 3$
(e.g.,
$\rho_3(1690)$),
in principle, can contribute. We make a natural assumption that the
heavy meson contributions are suppressed by small
$\sim \frac{1}{M}$
factors. Next, one can check that in the
$S=-1$
baryon sector only resonances with
$J=\frac{3}{2}, \; \frac{5}{2},...$
contribute to these sum rules. In this way we also manage to evade
the problem with poorly established spin
$\frac{1}{2}$
resonances over the
$\bar{K}N$
threshold. In our present analysis we are not going to take account of
possible contributions from exotic resonances with strangeness
 $S=+1$.
However, in what follows we show that several sum rules provide an
indirect evidences in favor of existence of exotics.

\begin{figure*}
\begin{center}
 \includegraphics[height=3.0cm]{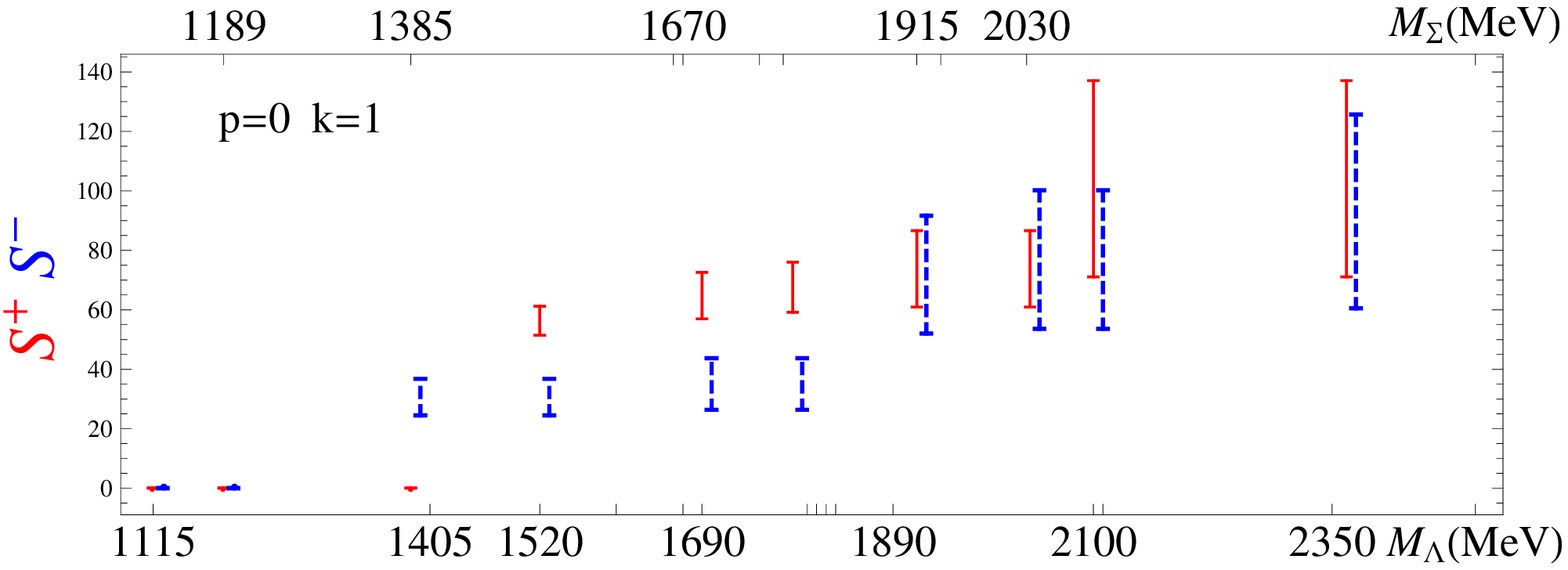}
  \includegraphics[height=3.0cm]{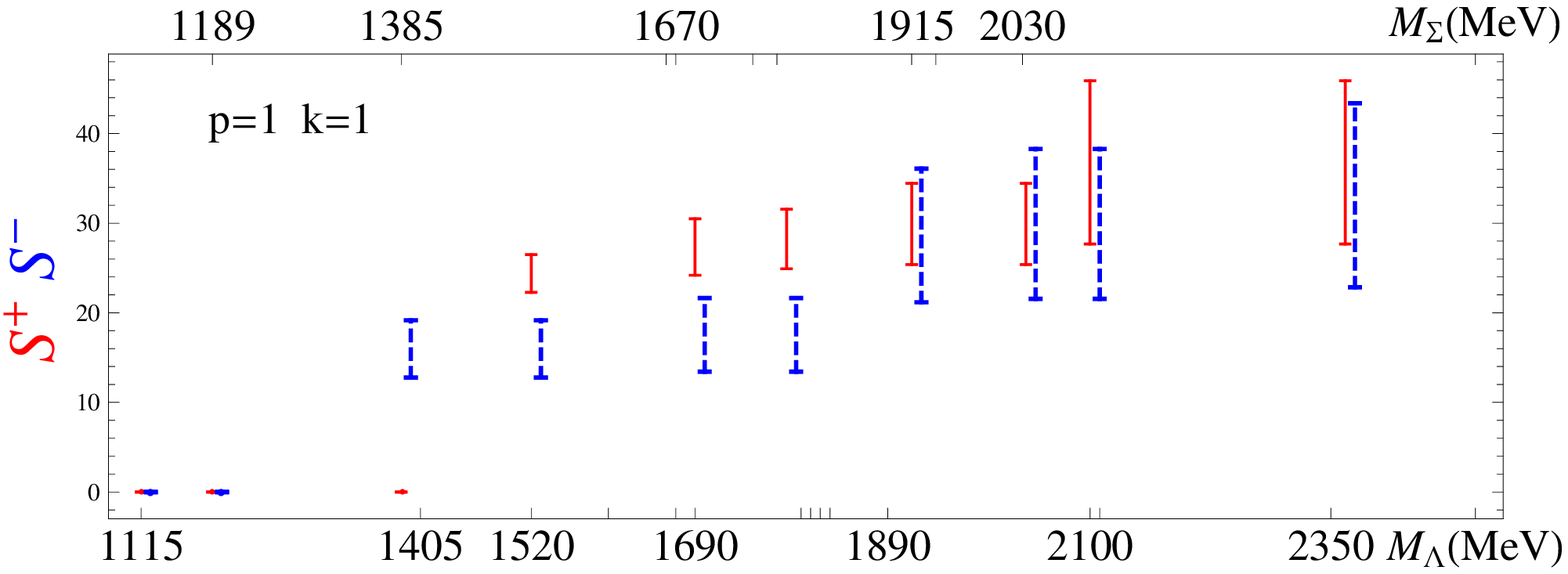}
   \includegraphics[height=3.0cm]{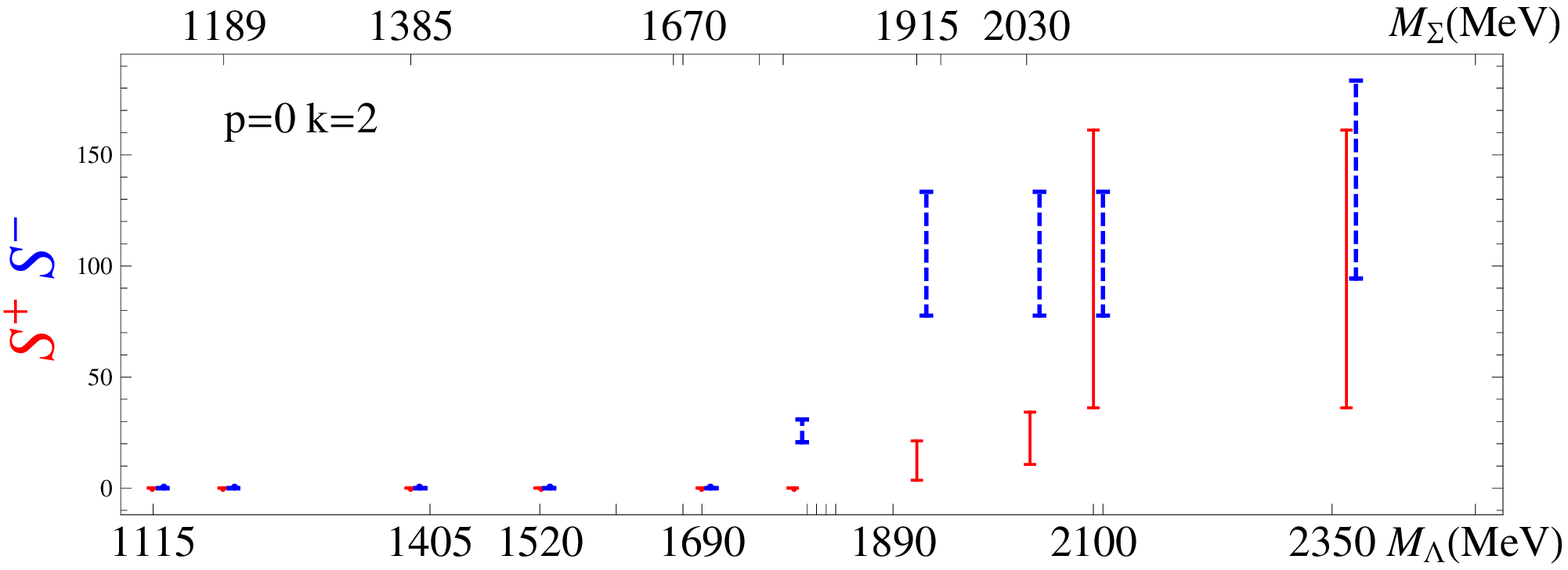}
    \includegraphics[height=3.0cm]{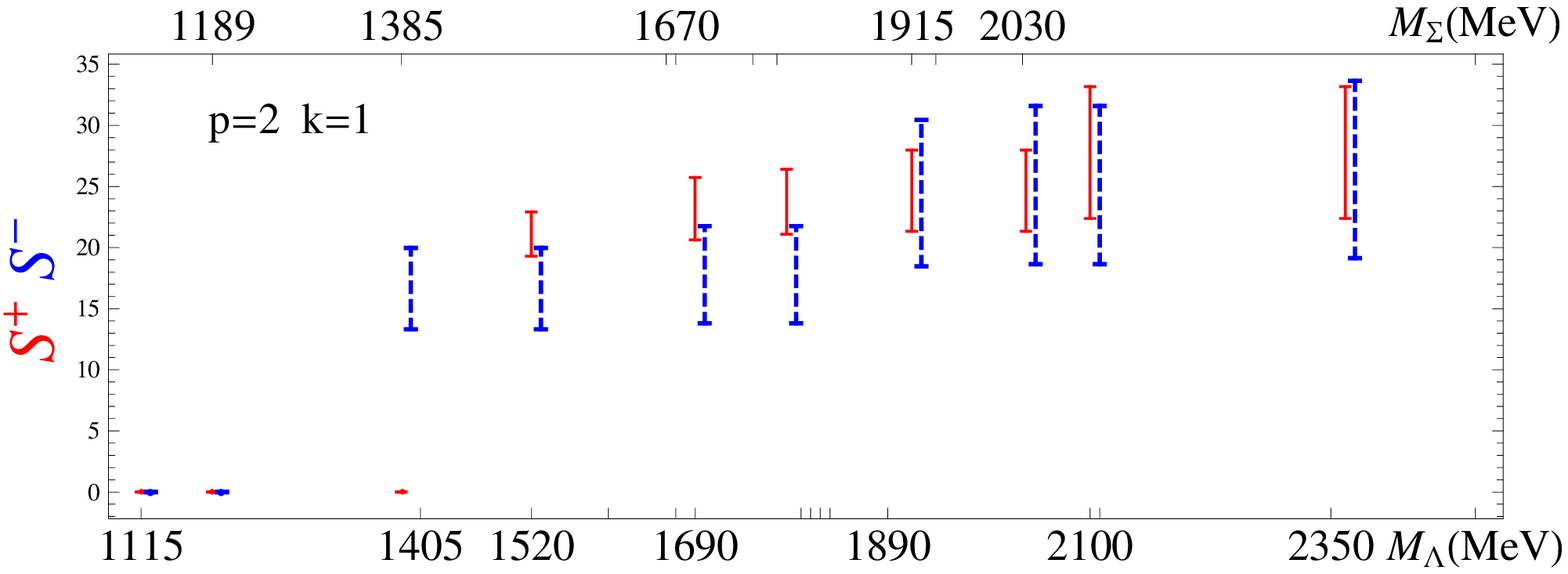}
     \includegraphics[height=3.0cm]{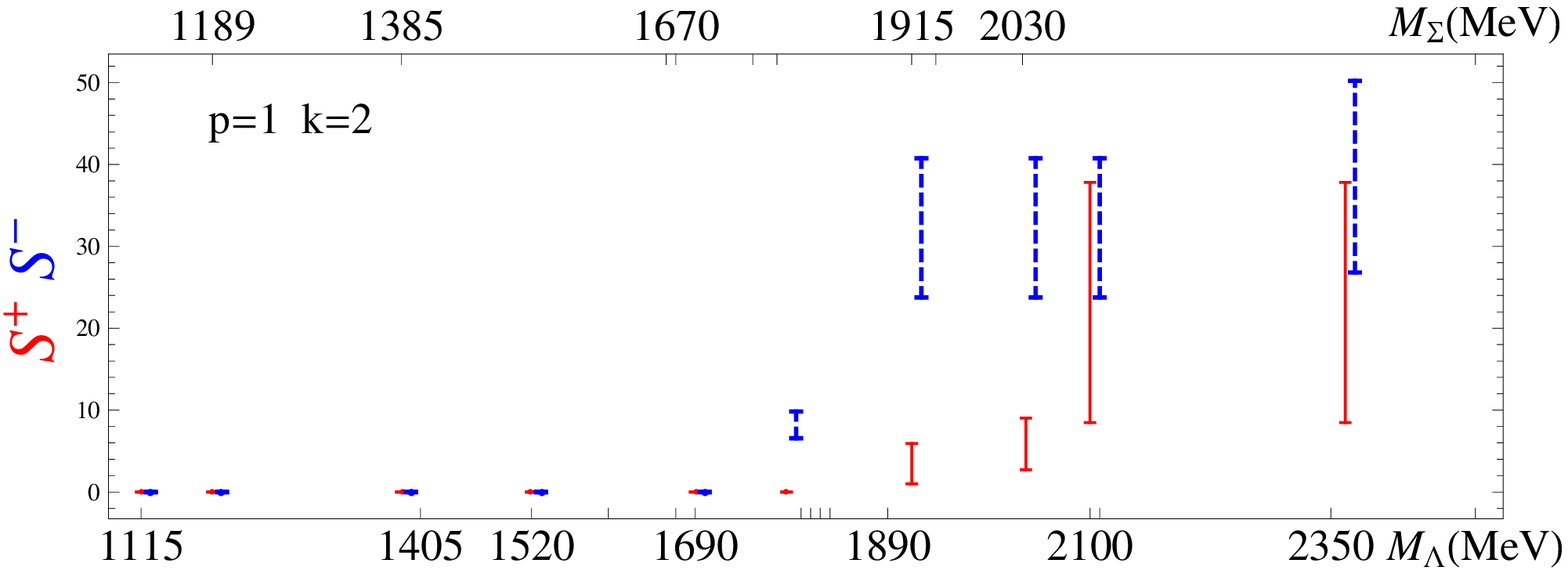}
      \includegraphics[height=3.0cm]{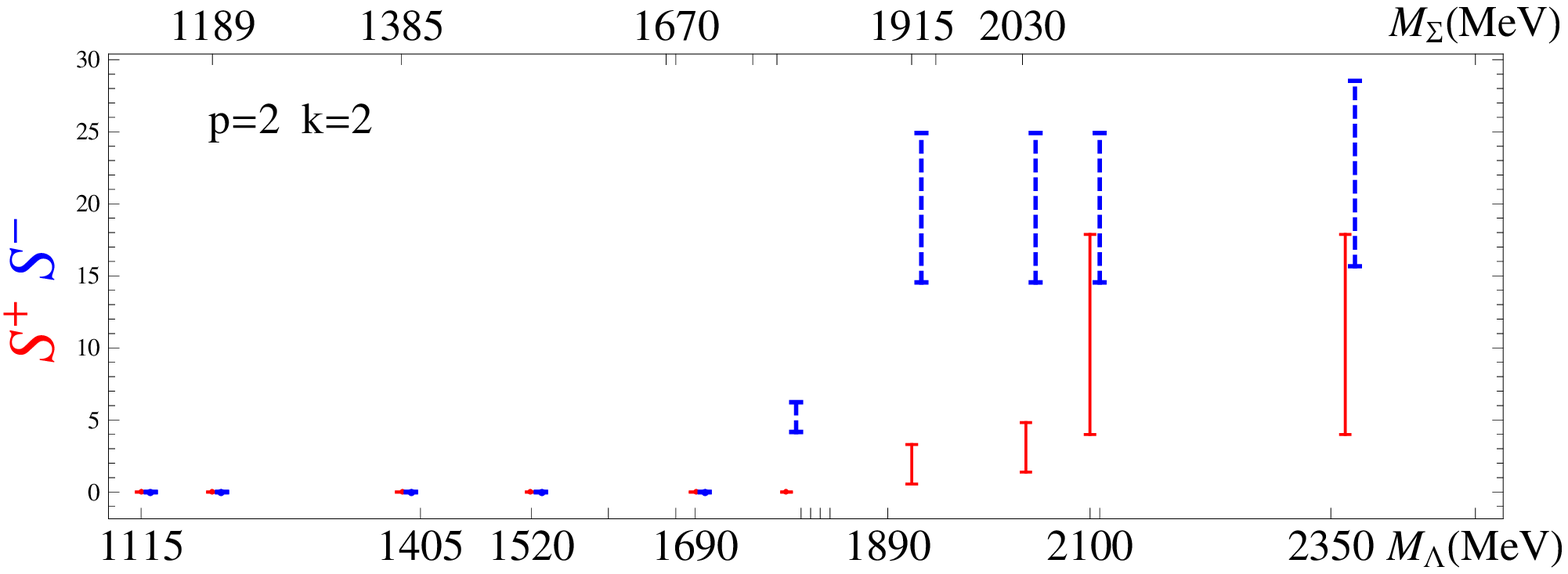}
  \caption{
Numerical tests of sum rules following from bootstrap constraints
(\ref{SR_KN_Aminus_US})
for different values of
$k$
and
$p$.
$S^+$
(solid) v.s.
$S^-$
(dashed) as functions of the heaviest
$S=-1$
baryon resonance taken into account. The error bars corresponding to
$S^-$
are shifted by
$10 \,  {\rm MeV}$
to the right from the resonance position for better discernibility. }
\label{KNbootPic1}
 \end{center}
\end{figure*}

To characterize the convergency of a given sum rule we introduce the
partial sums
$S^+(M_R)$
and
$S^-(M_R)$
of positive and negative contributions, respectively. For example, for
$\Phi_{A^-}(u,s)$
the partial sums are defined as:
\be
&&
S^+(M )= \sum_{{R_s \, R_t \, R_u,} \atop  M_R \le M}
(\partial_u)^p (\partial_s)^k \phi_{A^-}(u,s)
_{u=0 \atop s=0},  \nonumber \\ &&  \text{where every item}\ \ \
(\partial_u)^p (\partial_s)^k \phi_{A^-}(u,s)
_{u=0 \atop s=0} \ge 0;
\nonumber
\ee
\be
&&
S^-(M )= \sum_{{R_s \, R_t \, R_u,} \atop  M_R \le M}
 \left|
 (\partial_u)^p (\partial_s)^k \phi_{A^-}(u,s)
 \right|_{u=0 \atop s=0},  \nonumber \\ &&
  \text{where every item}\ \ \
 (\partial_u)^p (\partial_s)^k \phi_{A^-}(u,s)
_{u=0 \atop s=0}<0.
\nonumber
\ee
Here
$\phi_{A^-}$
stands for the individual resonance contribution to the generating
function
$\Phi_{A^-}(u,s)$.
Clearly, if
$S^+ \approx S^-$
the sum rule in question can be considered as well saturated. On
Figures~\ref{KNbootPic1},
\ref{New_SR},
\ref{BootFigKN4},
\ref{exoticPic}
for different sum rules we represent the dependence of the corresponding partial sums
$S^+$
and
$S^-$
on the mass of heaviest
$u$-channel
resonance taken into account. The error bars for
$S^+$
and
$S^-$
originate mainly from the uncertainties of the resonance decay widths
(and, hence, of triple coupling constants).

To make the domains of intersection of error bars of
$S^+$
and
$S^-$
better visible on our Figures
(Fig.~\ref{KNbootPic1}, \ref{New_SR}, \ref{BootFigKN4}, \ref{exoticPic})
the error bars corresponding to
$S^-$
are shifted by
$10 \, {\rm MeV}$
to the right from the true resonance position.

The results of saturation of several first sum rules that follow from
the bootstrap constraints
(\ref{SR_KN_Aminus_US})
are shown on
Fig.~\ref{KNbootPic1}.
One can conclude that these sum rules seem to be very well saturated
by known experimental data on the
$S=-1$
baryon resonances with
$J=\frac{3}{2}, \, \frac{5}{2}, \frac{7}{2}$
and masses
$< 2.4 \,  {\rm GeV}$.

The similar sets of well saturated sum rules also follow from the
bootstrap constraints for
$B^-$
invariant amplitude in the domain
$D_t$
(with generating function
$\Psi_{B^-}$)
and for
$A^-$
in
$D_s$
(with generating function
$\Psi_{A^-}$)
and
$D_u$
(with generating function
$\phi_{A^-}$).
There are also some other reasonably well saturated sum rules which we
do not show here. It seems highly improbable that the nice agreement
with data of the large number of sum rules stemming from bootstrap
constraints in three distantly lying domains of the Mandelstam plane
can be explained by accidental luck. So we conclude that the crucial
assumptions of our EST approach at least do not contradict roughly to
the known phenomenology of
$KN$
scattering.

\subsection{The importance of the subtraction terms}
\label{sub_sec_importance}

In order to stress the importance of the proper formulation of the
uniformity principle
(see Section~\ref{sec-Cforms})
and the necessity to take account of the correcting polynomials in
the Cauchy forms for invariant amplitudes in the hyperlayers with
non-decreasing asymptotic behavior let us perform the following
exercise. Consider the sum rules that follow from the bootstrap
constraint for the invariant amplitude
$A^-$
in the domain
$D_u$.
We are going to compare the saturation of the sum rules obtained
under the
{\it incorrect} assumption that
$A^-$
has decreasing asymptotic behavior in the hyperlayer
$B_t$
with the sum rules obtained under the correct asymptotic assumption
(in fact,
$A^-$
is 0-bounded function in the hyperlayer
$B_t$).

As an example we have chosen the simplest sum rule following from the
requirement that the Cauchy form for
$A^-$
in the hyperlayer
$B_t$
should coincide with that constructed in the hyperlayer
$B_s$
at the central point
$\{s=0,\,t=0\}$
of the domain
$D_u$:
\be
A^-(t=0,s=0)|_{B_t}=A^-(s=0,t=0)|_{B_s}\,.
\label{SR_without_corr_p}
\ee
Under the incorrect assumption on the asymptotic behavior of
$A^-$
in hyperlayer
$B_t$
the corresponding Cauchy form is just the sum over
$s$-
and
$u$-channel poles. On the contrary, under the proper asymptotic
assumption
(\ref{AB_minus_Bt_ass_cond})
it involves the correcting polynomials of
$0$th
degree in
$s-u$
as well as the smooth (background) term
$\alpha_{A^-}(t)$.
The background term can be computed from the bootstrap constraint in
the domain
$D_s$
with the help of the first kind bootstrap constraint
(\ref{alpha0-(t)}).
The sum rule in question follows from the bootstrap constraint of the
second kind
(\ref{concalpha0-})
with
$p=0$
and takes the form:
\be
\Psi_{A^-}(0,\,0)+\varphi_{A^-}(0,\,0)=0\,.
\label{SR_with_corr_p}
\ee

\begin{figure*}
\begin{center}
 \includegraphics[height=3.0cm]{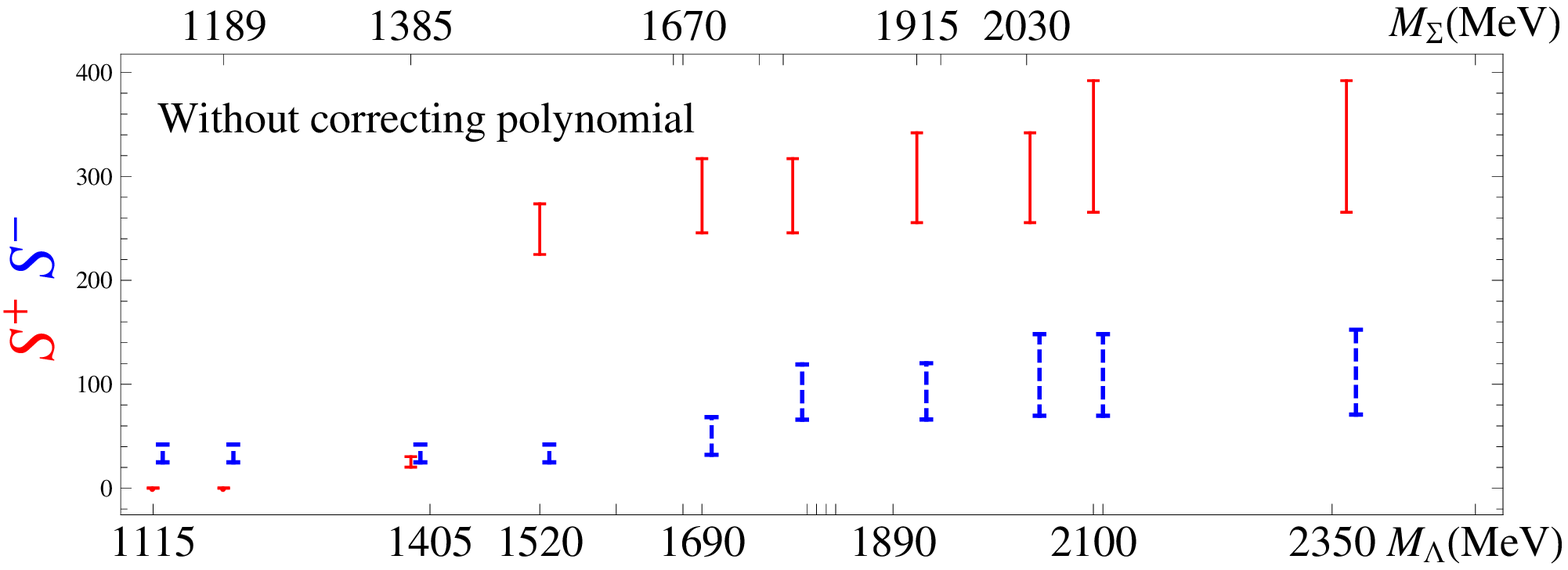}
   \includegraphics[height=3.0cm]{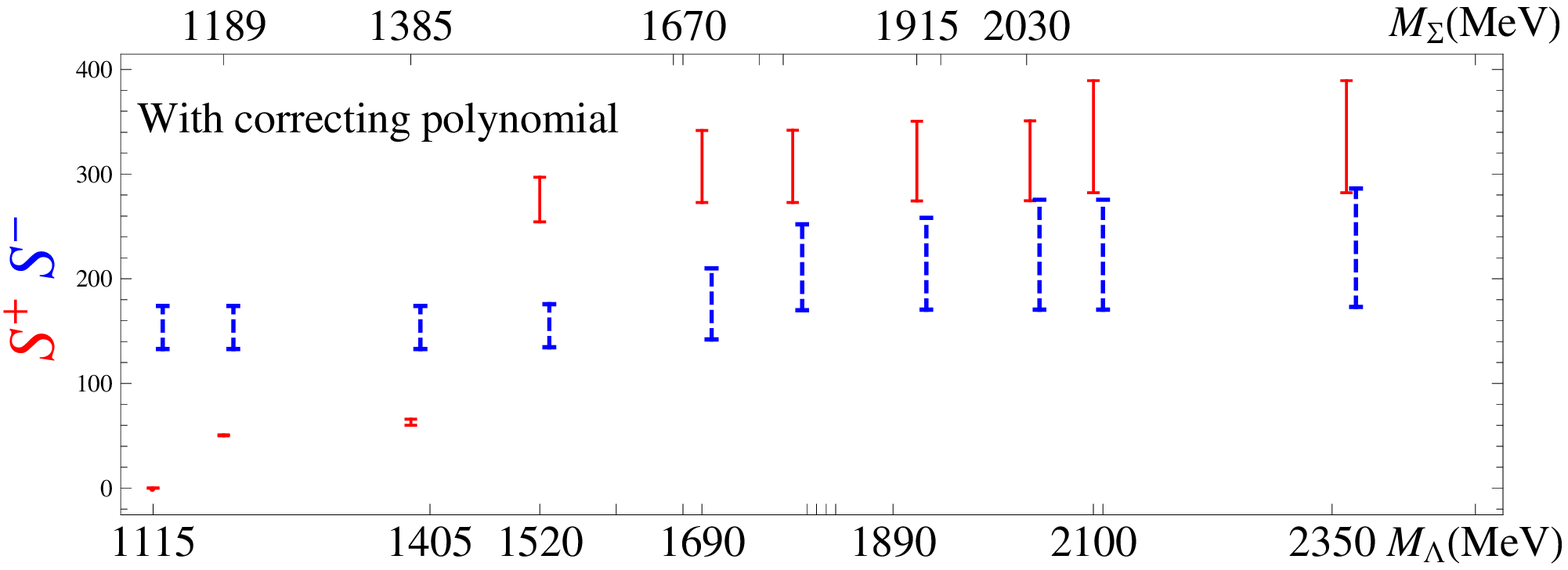}
\caption{Numerical tests of sum rule
for the invariant amplitude
$A^-$
in the domain
$D_s$.
On the upper panel we show the saturation of the sum rule
(\ref{SR_without_corr_p})
written without correcting polynomials and background term for
$A^-$
in
$B_t$.
On the lower panel we show the saturation of the sum rule
(\ref{SR_with_corr_p})
which takes account of the proper asymptotic condition for
$A^-$
in that hyperlayer.}
\label{New_SR}
 \end{center}
\end{figure*}

The results of saturation of the sum rule obtained under incorrect
assumption on the asymptotic behavior of
$A^-$
in
$B_t$
and the correct sum rule
(\ref{SR_with_corr_p})
with existing data are shown on
Fig.~\ref{New_SR}.
Note that this sum rule also receives contribution from the
$t$-channel
meson resonances with
$I=1$.
Because of poor knowledge of the relevant meson couplings we take
account of contribution of the lightest
$\rho(770)$
meson which is supposed to be dominant in the meson sector. The value
of the corresponding coupling
$G_1$
(see
\ref{Wa})
is taken from
\cite{Nagels,Dumbrajs:1983jd}.
It is clearly visible that the sum rule following from the correct
suggestion on the asymptotic behavior of
$A^-$
in the layer
$B_t$
is much better saturated than the sum rule without subtraction term
and correcting polynomials.

This example demonstrates the importance of both principles
(summability and uniformity) which give rise to the system of
bootstrap constraints in EST approach.
Under the incorrect assumption on the asymptotic behavior of invariant
amplitudes one could hardly expect to fulfil the analyticity
requirements encoded in the system of bootstrap conditions.

%%%%%%%%%%%%%%%%%%%%%%%%%%%%%%%%%%%%%%%%%%%%%%%%%%%%%%%%%%%%%%%%%%%%70

\subsection{On slowly converging sum rules}

In
Section~\ref{Sec-5.1}
we  presented an impressive series of well saturated sum rules for
$KN$
resonance parameters following from the system of bootstrap
conditions. However the situation with some other sum rules looks less
optimistic. Thus we have to perform the more detailed analysis. It
looks natural to discuss the possible reasons for which ceratin sum
rules cannot be saturated by the presently available data.

As an example, let us consider a particular sum rule which follows from
the second kind bootstrap constraints
(\ref{consPhiA+})
for the invariant amplitude
$A^+$
in
$D_t$:
\begin{equation}
(\partial_u)^p (\partial_s)^k
\Phi_{A^+}(u,s)
\Big|_{u=0 \atop s=0}=0,
\ \ \ {\rm for \ \ all } \ \
p,k =0,1,... \ \ .
\label{SR_KN_Plus_US}
\end{equation}

\begin{figure}
 \begin{center}
 \includegraphics[height=3.0cm]{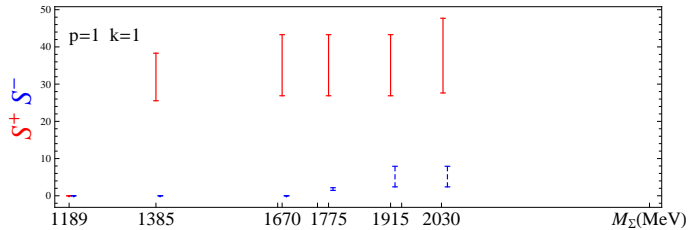}
\caption{
Numerical test of the sum rule from the system of constraints
(\ref{SR_KN_Plus_US})
for
$k=1$
and
$p=1$.
Partial sums
$ S^\pm$
are shown as functions of mass parameter of the heaviest
$\Sigma$
hyperon taken into account. }
\label{BootFigKN4}
 \end{center}
\end{figure}

The results of saturation of the sum rule that follows from
(\ref{SR_KN_Plus_US})
at
$p=k=1$
are shown on
Fig.~\ref{BootFigKN4}.
$\Lambda$
hyperons do not contribute and, at first glance, nothing can
compensate the huge positive contribution of
$(I=1, J=\frac{3}{2})$
resonances closest to threshold. Let us discuss the possible way to
overcome this difficulty.

First of all, we would like to recall that a similar situation was
encountered in the ``toy bootstrap model''
for the Lovelace string-like amplitude
\cite{POMI}.
To achieve the reasonable accuracy in the course of numerical testing
of certain sum rules for the resonance parameters in this model, it
was sufficient to take account of the contributions of relatively
small number of lowest poles. At the same time, when saturating some
other sum rules it was necessary to take account of the contributions
from considerable number of poles in one variable in order to
compensate the
``accidentally large''
contribution coming from just few first poles in another one.

The numerical testing of the toy bootstrap model has shown that
saturating of a given sum rule with finite number of lowest resonances
may result in imbalance due to several typical reasons:
\begin{enumerate}
\item
The sum rule belongs to the class of ``bad'' ones because it converges
very slowly. To achieve reasonable accuracy in the process of
numerical saturating, one has to take account of contributions from
the very large number of distant poles;
\item
It may turn out that one employs incorrect information on the
asymptotic behavior of the amplitude, so that the sum rule under
consideration is, in fact, divergent. This scenario was illustrated in
Section~\ref{sub_sec_importance};
\item
Finally, the information on the resonance spectrum may be incomplete
and certain light resonances, which might provide a considerable
contribution, are missed. If it is possible to point out the
resonance which helps to restore the balance in a given sum rule and,
at the same time, does not lead to problems with saturation of another
ones, this can be considered as  indirect evidence in favor of the
existence of such resonance.
\end{enumerate}
In what follows we are going to check whether some of these scenarios
can be applied to the sum rules
(\ref{SR_KN_Plus_US}).

\begin{figure}
\begin{center}
\includegraphics[height=2.5cm]{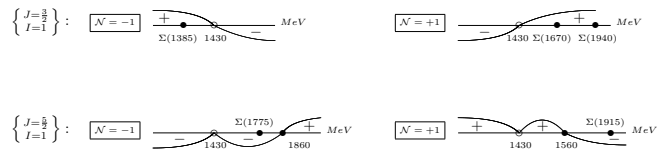}
\caption{
Signs of the contributions from
$\Sigma$
hyperons with spin
$j=\frac{3}{2}, \,\frac{5}{2}$
and normality
$\mathcal{N}=\pm 1$
to the sum rule following from bootstrap constraint
(\ref{SR_KN_Plus_US})
with
$p=k=1$.}
\label{BootFigKN3}
\end{center}
\end{figure}

The first scenario looks promising. One can easily specify the quantum
numbers of heavy baryon resonances whose contributions to the sum rule
(\ref{SR_KN_Plus_US})
with
$k=p=1$
enter with suitable signs. On
Figure~\ref{BootFigKN3}
we show the behavior of signs of contributions to the sum rule
(\ref{SR_KN_Plus_US})
with
$p=1, \, k=1$
from different
$S=-1$
$\Sigma$
hyperon families. We conclude that the contributions of the
``tail''
of heavy resonances with
$ J=\frac{3}{2}, \, \mathcal{N}=-1$
and
$ J=\frac{5}{2}, \, \mathcal{N}=+1$
may gradually compensate the large contribution from
$\Sigma(1385)$.
The same mechanism then also works for other sum rules from this group
with
$k>1$.

The Regge theory intercepts are known since 1960's and are much
reliable. Thus, at first glance, the second of the above-mentioned
scenarios seems to be of little use. Nevertheless, let us stress that
the additional information on the high energy asymptotic behavior of
invariant amplitudes of binary scattering processes at various fixed
values of other kinematical variables is
{\em highly demanded}.
The thing is that the sum rules following from the bootstrap
constraints turn out to be slowly converging in the vicinities of
domains where the asymptotic regime changes. These sum rules require
application of methods of convergency acceleration.

Now we are going to discuss whether we may profit from the possibility
that certain resonances which could give a significant contribution to
our sum rules have been for some reasons omitted.

The sum rule
(\ref{SR_KN_Plus_US})
with
$p=1, \, k=1$
may be considered as a purely baryonic one: in the meson sector only
resonances with spin
$j \ge 3$
may contribute. Thus to saturate it is natural to look for possible
amendments to
$KN$
spectrum in the baryon sector. Indeed the ``world average'' PDG
$KN$
resonance spectrum
\cite{PDG}
may be incomplete or imprecise both in the
$u$-, $t$-
and
$s$-channel
sectors of
$KN$
reaction. We would like to recall (see
\cite{KSAVVV3})
that the numerical test of
$\pi N$
bootstrap sum rules employing the more precise
$\pi N$
spectrum obtained with the help of advanced coupled-channel partial
wave analysis
\cite{Arndt:2006bf}
rather than PDG data
\cite{PDG}
shows considerable improvement of saturation results. We strongly
suggest the application of methods employed in
\cite{Arndt:2006bf}
for the
$KN$
elastic scattering. A refined spectrum of
$\Lambda$
and
$\Sigma$
hyperons would allow the high precision tests of
$KN$
bootstrap sum rules. It would be also extremely interesting to try to
take account of the bootstrap constrains for resonance parameters at
the level of partial wave analysis. On the other hand the problem of
theoretical development of fitting procedures directly based on the
effective scattering theory approach (in which the notion of
resonance is rigorously defined) also awaits its solution.

Another possibility to amend the
$KN$
spectrum is to assume the existence of resonances (so called
$Z$
baryons) in the
$s$-channel.
Up to present time such a possibility has not been excluded
by experiment. One of the most tempting candidates is the exotic
baryon resonance with strangeness
$S=+1$ --
so called
$\theta(1530)$.
Since the prediction of its mass and width in
\cite{Diakonov}
and the first experimental publications
\cite{Nakano,Barmin}
there was much interest to light and narrow exotic resonances. So far
experiment does not show a clear-cut evidence of their existence. For
the discussion and review of experimental situation see
\cite{Review_theta-1,KenHicks,Burkert,Review_theta-2}.
It is very interesting to see whether our sum rules can be of any use
for clarifying this issue.

For example, one can try to interpret the deficit in
(\ref{SR_KN_Plus_US})
with
$p=1, \, k=1$
(as well as in some other sum rules) as indirect evidence of the
existence of exotic baryon (or few such baryons). To analyze this
possibility in more detail we need to discuss first the
characteristics (quantum numbers and widths) of the exotic resonances
which could manifest themselves in our sum rules.

Unfortunately the ordinary
spin-$\frac{1}{2}$
narrow exotic resonance cannot make significant contribution to our
sum rules. Several recent results of data analysis estimate the
possible
$\theta$
decay width as
$\Gamma_{\theta  \rightarrow KN} \, \sim  \, 1 \, {\rm MeV}$.
This is consistent with the advanced theoretical estimates of
$\theta$
width in the framework of Chiral Quark Soliton model
\cite{DPPcomments, Lorce:2006nq}.
This makes the coupling  of
$\theta$
resonance to
$KN$
extremely small. Its possible contribution is invisible against that
of background from poorly established
$\Lambda$
and
$\Sigma$
hyperon resonances. This situation is quite similar to that with
the contribution of recently established (see
\cite{Oller})
narrow resonances with
$S=-1$.

Thus we conclude that the possibility to get
information on
spin-$\frac{1}{2}$
narrow exotic resonance from our sum rules looks unrealistic. However,
the possibility that exotic resonance can be a higher spin state still
is not excluded
\cite{Nishikawa:2004tk,Nam:2005jz,Kim,Gubler:2009iq}.

\begin{figure}%[ht]
\begin{center}
\includegraphics[height=3cm]{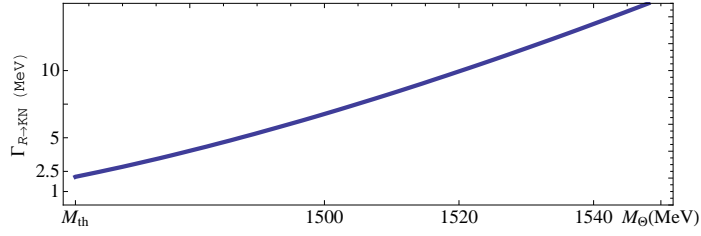}
    \caption{The parametric dependence of the decay width of
  $J^P={\frac{3}{2}}^-$
  exotic resonance on its mass for the fixed value of $G_{KN \theta}=25$.}
\label{exoticPic}
\end{center}
\end{figure}

Among the possible choices of quantum numbers of exotic resonance
$\theta$
$J^{\mathbf{P}}={\frac{3}{2}}^-$
has the advantage that the small width of resonance is quite
compatible with its significant contribution to our sum rules. On
Figure~\ref{exoticPic}
we show the parametric dependence of the exotic
$\theta$ $J^{\mathbf{P}}={\frac{3}{2}}^-$
decay width on the resonance mass for the fixed value of dimensionless
coupling constant
$G_{KN \theta}=25$
(see (\ref{GKNR})).
Below we show that this value of
$G_{KN \theta}$
allows one to saturate the sum rule
(\ref{SR_KN_Plus_US})
with
$p=1, \, k=1$
by the contribution of exotics.

We conclude that for a
$J^{\mathbf{P}}={\frac{3}{2}}^-$
resonance with the mass parameter
$M \sim 1530 \; MeV$
the decay width which corresponds to
$G_{KNR}=25$
is about
$10 \, MeV$.
It can be reduced to several
$\rm MeV$
by shifting the mass parameter towards $KN$ threshold value.
By the way, one can check that for the resonance with spin
$J^{\mathbf{P}}={\frac{3}{2}}^+$
and mass
$\sim 1530 \, MeV$
the decay width that corresponds to
$G_{KNR}=25$
is
$\sim 500 \, MeV$.
That is why this choice of quantum numbers seems less favorable in our
approach if we suppose the exotic resonances to be narrow and, at the
same time, providing sizeable contributions to sum rules.

\begin{figure}%[ht]
\begin{center}
\includegraphics[height=3cm]{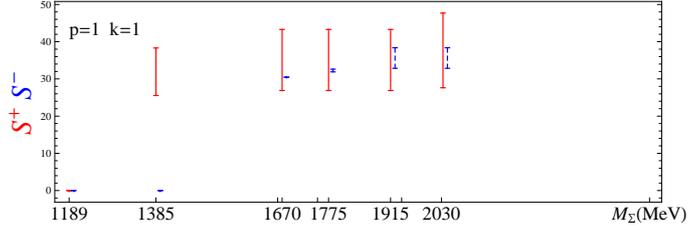}
    \caption{Numerical test of sum rule
  following from the bootstrap constraint
  (\ref{SR_KN_Plus_US})
  with
  $k=p=1$
  taking into account the contribution of the exotic
  $S=+1$
  resonance in the $s$-channel with
  $J^{\mathbf{P}}={\frac{3}{2}}^-$ with $M=1530 \; {\rm MeV}$ and
  $\Gamma_{KN} \sim 10 \; {\rm MeV}$
  Compare to
  Fig.~\ref{BootFigKN4}. }
\label{exoticPicRes}
\end{center}
\end{figure}

On
Fig.~\ref{exoticPicRes}
we show that the contribution of the exotic baryon state with
$J^{\mathbf{P}}={\frac{3}{2}}^-$
(normality
$\mathcal{N}=+1$)
above the
$KN$
threshold could significantly compensate the deficit in the bootstrap
constraint
(\ref{SR_KN_Plus_US})
with
$k=p=1$.
It can be shown that such a resonance could also improve the result of
saturation of several other sum rules without breaking the balance in
those relations which have been earlier saturated without attraction
of
$S=+1$
resonances.

In fact this exercise only gives an idea how the bootstrap constrains
can be used for the needs of hadron spectroscopy. Obviously, for the
moment we still lack the precise information on
$u$-
and
$t$- channel
$KN$
resonance spectrum in order to undertake the search for
``realistically narrow''
($\Gamma < 1\, {\rm MeV}$)
exotic
$KN$
resonances.

\section{Conclusion}
\label{conclusion}
%\mbox{}

We argue that the concept of
the effective scattering
theory may be of considerable
practical use for understanding the strong processes. %In
%\cite{AVVV2} -- \cite{KSAVVV2}
%we have formulated rules needed to handle an effective scattering
%theory of strong interactions in the
%$S$-matrix sector.
The requirement of existence of
%possibility to construct
a rigorously defined Dyson perturbation expansion at every fixed loop order was
used in \cite{AVVV2} -- \cite{KSAVVV2}
as a key to solve  multiple specific problems  that emerge
when dealing with an infinite component effective theory in the
$S$-matrix sector.
%It was shown that the
Fundamental requirements of covariance, unitarity,
causality and crossing together with
assumption on realistic
asymptotic behavior of invariant amplitudes result in a set of
{\em bootstrap conditions}
for the physical (measurable) parameters of effective Hamiltonian. The
remarkable property of
{\em renormalization invariance}
of the system of bootstrap conditions makes it possible the direct
comparison of sum rules following from this system with experimental
data.

In this paper we apply our general EST scheme to the
description of hadron binary scattering in the strange sector. We
construct the well-defined tree level amplitudes
of
$KN$
scattering in three intersecting layers
$B_s$, $B_t$, $B_u$
and  derive the system of bootstrap conditions for
$KN$
resonance parameters. The numerical tests of corresponding sum rules
make it possible to claim the consistency of our approach with
presently known phenomenology. The additional arguments in favor of
this statement will be given in a special publication on
mathematical aspects of numerical testing the bootstrap conditions.

We also show that the sum rules that follow from the system of bootstrap
conditions can be used as a tool to study the hadron spectrum and bring
indirect evidence in favor of existence of
exotic resonances in the
$s$-channel
of elastic
$KN$
scattering.

%%%%%%%%%%%%%%%%%%%%%%%%%%%%%%%%%%%%%%%%%%%%%%%%%%%%%%%%%%%%%%%%%%%%70

\section*{Acknowledgements}

We are grateful to A.~Vereshagin for useful discussions, valuable
help and advises. Also we thank M.~Braun, V.~Franke, S.~Paston,
I.~Strakovsky, A.~Tochin, and M.~Vyazovsky for multiple discussions
on various aspects of effective scattering theory approach and its
application to hadron spectroscopy.

%%%%%%%%%%%%%%%%%%%%%%%%%%%%%%%%%%%%%%%%%%%%%%%%%%%%%%%%%%%%%%%%%%%%70

\appendix
\section{Miscellaneous}
\label{app1}

\subsection{Kaon and Nucleon field parametrization}
\label{appA2}

Kaon (nucleon) fields are parameterized with the help of covariant and
contravariant isotopic spinors
$|K_i \rangle, \; |\overline{K}^i \rangle$
($|N_\alpha \rangle, \; |\overline{N}^\alpha \rangle$)
which transform under isotopic transformations
as follows:
\be
 I_a  |K_i \rangle =
 \left( \sigma_a/2  \right)^{j \, \cdot}_{\cdot \, i} |K_j
 \rangle\,; \ \ \ \
 I_a  |\overline{K}^i \rangle =
 - \left( \sigma_a/2 \right)^{i\, \cdot}_{\cdot \, j}
 |\overline{K}^j \rangle.
\nonumber
\label{contrv}
\ee
Here
$i,j,\alpha= \{1,2 \}$
stand for the isotopic spinor indices and
$\sigma_a$ --
for the Pauli matrices:
$Tr(\sigma_a \sigma_b)=2 \delta_{ab} \ \ (a,b= \{1, \,2,\,3 \})$.

The list the isotopic projecting operators for three channels of
$KN$
elastic scattering reaction looks as follows:
\begin{itemize}
\item
$s$-channel:
\be
&& {P^{s(0,\,1)}}^{ \cdot \, \cdot \, \beta j}_{ \alpha i \, \cdot \, \cdot}=
\frac{\delta_{\alpha \, \cdot}^{\cdot \, \beta } \delta_{i \, \cdot}^{\cdot \, j} \mp
\delta_{\alpha \, \cdot}^{\cdot \, j} \delta_{i\,\cdot}^{\cdot \, \beta}}{2};
\label{prjS}
\ee
\item
$t$-channel:
\begin{equation}
{P^{t \, (0) }}^{\cdot \, \beta \, j \, \cdot}_{\alpha \, \cdot \; \cdot \,  i}=
%P^{t(0) \;  \beta j}_{\ \  \alpha \ \ \; i}=
\frac{\delta_{\alpha\, \cdot}^{\cdot \, \beta } \delta_{i\, \cdot}^{\cdot \, j}}{2}, \ \ \ \
%P^{t(1) \;  \beta j}_{\ \  \alpha \ \ \; i}=
{P^{t \, (1) }}^{\cdot \, \beta \, j \, \cdot}_{\alpha \, \cdot \; \cdot \,  i}=
\delta_{\alpha \, \cdot}^{\cdot \, j} \delta_{i \, \cdot}^{\cdot \, \beta}-
\frac{\delta_{\alpha \, \cdot}^{\cdot \, \beta } \delta_{i\, \cdot}^{\cdot \,
j}}{2} \;;
\label{prjT}
\end{equation}
\item
$u$-channel:
\begin{equation}
{P^{u \, (0) }}^{\cdot \, j \, \beta \, \cdot}_{\alpha \, \cdot \; \cdot \,  i}=
%P^{u(0) \;  j \beta  }_{\ \  \alpha \ \ \; i}=
\frac{\delta_{\alpha \, \cdot} ^{\cdot \, j } \delta_{i \cdot}^{\cdot \,
\beta}}{2}, \ \ \ \
{P^{u \, (1) }}^{\cdot \, j \, \beta \, \cdot}_{\alpha \, \cdot \; \cdot \,  i}=
\delta_{\alpha\, \cdot}^{\cdot \, \beta} \delta_{i \, \cdot}^{\cdot \, j}-
\frac{\delta_{\alpha \, \cdot}^{\cdot \, j} \delta_{i \, \cdot}^{\cdot \,
\beta}}{2} \;.
\label{prjU}
\end{equation}
\end{itemize}

Let us also explain the isospin parametrization of resonance fields.
For instance, if the resonance has isospin
$I = 1$
the corresponding field operator reads
\[
{P^{u \, (1) }}^{\cdot \, j \, \beta \, \cdot}_{\alpha \, \cdot \; \cdot \,  i} R_{\beta\, \cdot}^{\cdot \, i}
\equiv {R^{(1)}}_{\alpha \, \cdot}^{\cdot \, j}= {(\sigma_a)}_{\alpha \, \cdot}^{\cdot \, j} R^a
%P^{(1) \;  j k}_{\ \  i \ \ \; l}
%R^{l}_{\; k}
%\equiv
%R^{(1)  j}_{i}
%= (\sigma_a)^i_j R_a, \ \ \ i,j,k,l=1,2; \ \ \ a=1,2,3.
\]
Thus, for isospin
$1$
one may use either one isovector index
$a=1,2,3$ or spinor notations with two spinor indices. The superscript
``$(1)$''
then implies the contraction with relevant isotopic projector.

\subsection{Asymptotic conditions}
\label{appA3}

Below we present a summary of Regge theory prescriptions for the
asymptotic behavior of the invariant amplitudes
$A^\pm,\ \ B^\pm$
that appear in
\Ref{invAml}.
This behavior is determined by the known intercepts
$a_I(0)$
of the leading Regge trajectories (see, e.g.,
\cite{Collins1977})
with the cross-channel isospin
$I$:
\be
&&
a_0(s)|_{s=0}<0, \quad a_1(s)|_{s=0}<0;
\nonumber \\ &&
 a_0(t)|_{t=0}=1, \quad 0<a_1(t)|_{t=0}<1;
 \nonumber \\ &&
a_0(u)|_{u=0}=-0.7, \quad a_1(u)|_{u=0}=-0.3.
\nonumber
\ee
In the boxes below we show the degrees of bounding polynomials needed
to construct the corresponding Cauchy forms in various layers.
\begin{itemize}
\item
$B_t: \{ \nu_t \in \mathbb{C}, \;  t \sim 0 \}$:
\be
&&
\left. (2A^++A^-)\right|_{|\nu_t| \to \infty} {\sim  }
o(|\nu_t|^2 )\,; \ \ \boxed{N_t(2A^++A^-)= 1}
\nonumber \\ &&
\left. (2B^++B^-)\right|_{|\nu_t| \to \infty} {\sim  } o(|\nu_t|)\,;   \ \ \  \boxed{N_t(2B^++B^-)=0}
\nonumber \\ &&
\ee
\be
&&
\left.  A^- \right|_{|\nu_t| \to \infty} {\sim  }
o(|\nu_t| )\,; \ \ \boxed{N_t(A^-)=0}
\nonumber \\ &&
\left.  B^- \right|_{|\nu_t| \to \infty} {\sim  } o(1)\,;  \ \ \ \ \, \boxed{N_t(B^-)=-1}
\label{AB_minus_Bt_ass_cond}
\ee
\item
In two remaining layers
($B_s$ and $B_u$)
all invariant amplitudes possess the decreasing asymptotic behavior
and neither correcting polynomials nor smooth terms are needed.
\end{itemize}

\section{Compact notations}
\label{app2}

In this Appendix we summarize the notations used in main text to
keep the results in a compact form. To describe the residues of the
invariant amplitudes at poles corresponding to baryon resonance
exchanges it is convenient to introduce two families of functions:
\be
%&&
F_{\sss A}^l(M, \chi)=
(M+m)P'_{l+1}(1+\frac{\chi}{2 \phi})
%\nonumber \\ &&
+(M-m)
\frac{(M+m)^2-\mu^2}{(M-m)^2-\mu^2}P'_l(1+\frac{\chi}{2 \phi})\,
\label{Fa}
\ee
and
\begin{equation}
F_{\sss B}^l(M, \chi)=P'_{l+1}(1+\frac{\chi}{2 \phi})-
\frac{(M+m)^2-\mu^2}{(M-m)^2-\mu^2}P'_l(1+\frac{\chi}{2 \phi})\, .
\label{Fb}
\end{equation}
$P_l$ stand here for the Legendre polynomials and
$\phi$ is the universal K\"allen function:
\be
%&&
\phi \equiv |\vec{k}|_{\sss C.M.F.}^2 %\nonumber \\ &&
 =
\frac{1}{4M^2}\, (M^4+m^4+\mu^4-2M^2m^2-2M^2 \mu^2-2m^2 \mu^2)\, .
%\nonumber \\
\label{phiK}
\ee

The residues of the invariant amplitudes
$(X=A,B)$
at poles corresponding  to the exchange with
$s$-
or
$u$-channel baryon resonance
with mass parameter
$M$,
spin
$j=l+\frac{1}{2}$
and normality
$\mathcal{N}$
are given (up to isotopical factors) by the expression:
\be
Y_{\sss X}(j=l+\frac{1}{2},\mathcal{N},M, \chi) =
G_{\sss KNR}F^l_X(-\mathcal{N}M,\chi),
\label{Y_X}
\ee
where $G_{\sss KNR}$ is a dimensionless constant
\be
G_{\sss KNR}= g_{\sss R} ^2 \frac{l!}{(2l+1)!!} \, \phi^l\,.
\label{dim_constant}
\ee

The residues of the invariant amplitudes at poles corresponding
to $t$-channel meson resonance exchanges with mass parameter
$M$
and spin
$l$
are given by the functions
$W_{\sss X}(M,l,\chi)$:
\be
&&
W_{\sss A}(M,l,\chi) = G_{\sss 1} P_l(\chi)-\frac{m}{m^2-
\frac{M^2}{4}} G_{\sss 2}P'_{l-1}(\chi), \nonumber \\ &&
W_{\sss B}(M,l,\chi)=\frac{1}{F}G_{\sss 2}P'_l(\chi),
\label{WaWb}
\ee
where
\be
F=\frac{1}{2} \sqrt{|(M^2-4m^2)(M^2-4\mu^2)|}\, ;\ \ \ \ \ \ \
\ee
and
\be
G_{\sss 1,2} =
g_{\sss {K\overline{K}R}} \, g^{\sss (1,2)}_{\sss {N\overline{N}R}}
\frac{j!}{(2j-1)!!} F^j\, .
\label{Wa}
\ee

To shorten our notations we introduce the sign factor
$\eta_{\sss X}$:
\be
\eta_{\sss X}= \left\{ {+1, \ \ X=A} \atop {-1, \ \ X=B} \right.
\label{numcoef1}
\ee
and  three sets of isotopic
($I=0,\,1$)
coefficients:
$b_{\sss I}^\pm$ --
for baryons with strangeness
$S=-1$,
$c_{\sss I}^\pm$ --
for hypothetic exotic baryons with strangeness
$S=+1$,
and
$ d_{\sss I}^\pm $ --
for non-strange mesons:
\begin{equation}
\begin{split}
&b_{\sss 0}^+= 0,\ \ \          b_{\sss 0}^- =+\frac{1}{2},
\ \ \
b_{\sss 1}^+ = 1,\ \ \           b_{\sss 1}^- =-\frac{1}{2}\,;  \\ &
c_{\sss 0}^+ =+\frac{1}{2},\ \ \  c_{\sss 0}^- =-\frac{1}{2}, \ \ \
c_{\sss 1}^+ =+\frac{1}{2},\ \ \  c_{\sss 1}^- =+\frac{1}{2}\,;\ \ \ \ \\ &
%,\ \ \ \
d_{\sss 0}^+ =+\frac{1}{2},\ \ \  d_{\sss 0}^- = 0,\ \ \
d_{\sss 1}^+ =-\frac{1}{2},\ \ \  d_{\sss 1}^- =1\, .
\label{numcoef2}
\end{split}
\end{equation}

\section{The explicit expressions for generating functions}
\label{app3}
\mbox{}

Here we give the explicit expressions for the generating functions of
the system of bootstrap constraints.
\be
\Phi_{\sss X^\pm} (u,s)  \nonumber  &&   =    \sum_{\sss B\{S=+1\}}
c_I^\pm Y_{\sss X}\left( ...,-(\Sigma+u) \right)
\frac{1}{s-M^2} - \sum_{\sss B\{S=-1\}}\eta_{\sss X} b_I^\pm
Y_{\sss X}\left( ...,-(\Sigma+s) \right)
\frac{1}{u-M^2}  %\notag
\nonumber \\ &&
+ \sum_{\sss M} \frac{d_I^\pm}{s+u+\Sigma}
\left\{ W_{\sss X}\left( ...,\frac{2s+\Sigma}{4F}\right)
\right.-
\left. W_{\sss X}\left( ...,\frac{-(2u+\Sigma)}{4F}\right)
\right\} .
\label{Phi-X-pm}
\ee
%\end{align}
%**************************************
\be
%\begin{align}
\varphi_{\sss A^+}(s,t)  \nonumber && =
 -\sum_{\sss B\{S=+1\}} Y_{\sss A}(...,t)
\left( \frac{c_I^+}{s-M^2}+\frac{2c_I^+}{t+2 \theta}+
\frac{2(c_I^+ +\frac{1}{2}c_I^-)(t+2s-2 \sigma)}{(t+2 \theta)^2}
\right)
\nonumber \\ && -
\sum_{\sss B\{S=-1\}}   \Bigg\{ Y_{\sss A}(...,t) \left(
\frac{-b_I^+}{s+t+ \Sigma}+\frac{2b_I^+}{t+2 \theta}-\frac{2(b_I^+
+ \frac{1}{2}b_I^-)(t+2s-2\sigma)}{(t+2 \theta)^2} \right)
\nonumber \\ && -
Y_{\sss A} \bigl(...,-(s + \Sigma)\bigr) \frac{-b_I^+}{s+t+ \Sigma}
\Bigg\} +
\sum_{\sss M}\frac{d_I^+}{t-M^2}\,
W_{\sss A}\left( ...,\frac{2s+\Sigma}{4F}\right).
\label{varphi-A+}
\ee
%\end{align}
%**************************************
%\begin{align}
\be
\Psi_{\sss A^+}(t,u)  = &&
\sum_{\sss B\{S=-1\}}  Y_{\sss A}(...,t)
\left(
\frac{b_I^+}{u-M^2}+\frac{2b_I^+}{t+2 \theta}-
\frac{2(b_I^++\frac{1}{2}b_I^-)(-t-2u+2 \sigma)}{(t+2 \theta)^2}
\right)
\nonumber \\ && +
\sum_{\sss B\{S=+1\}} \Bigg\{ Y_{\sss A}(...,t ) \left(
\frac{-c_I^+ }{t+u+ \Sigma} + \frac{2c_I^+ }{t+2 \theta}+
\frac{2(c_I^++\frac{1}{2}c_I^-)(-t-2u+2 \sigma)}{(t+2 \theta)^2}
\right) \nonumber \\ &&  %\ \ \ \ \ \
- Y_{\sss A}\bigl( ...,-(\Sigma + u) \bigr) \frac{-c_I^+}{t+u+\Sigma}
\Bigg\}
- \sum_{\sss M}\frac{d_I^+}{t-M^2}\,
W_{\sss A}\left(...,\frac{-(2u+\Sigma)}{4F}\right) .
\label{Psi-A+}
\ee
%**************************************

\be
\varphi_{\sss B^+}(s,t)  = &&
\sum_{\sss B\{S=-1\}}   \Bigg\{ Y_{\sss B}(...,t)
\left(\frac{-b_I^+}{s+t+ \Sigma} +
\frac{2(b_I^+ + \frac{1}{2}b_I^-)}{t+2 \theta} \right)
- Y_{\sss B}\bigl( ...,-(s+\Sigma) \bigr)
\frac{-b_I^+}{s+t+ \Sigma} \Bigg\}  \nonumber \\ && -
\sum_{\sss B\{S=+1\}} Y_{\sss B}(...,t) \left(
\frac{c_I^+}{s-M^2} + \frac{2(c_I^+ +\frac{1}{2}c_I^-)}{t+2 \theta}
\right)+ \sum_{\sss M} \frac{d_I^+}{t-M^2}
W_{\sss B}\left( ...,\frac{2s+\Sigma}{4F} \right) .
\nonumber \\ &&
\label{varphi-B+}
\ee
%\end{align}
%**************************************
%\begin{align}
\be
\Psi_{\sss B^+}(t,u)= &&
-\sum_{\sss B\{S=+1\}}\left\{
 \left. Y_{\sss B}(...,t)
\left( \frac{-c_I^+ }{t+u+ \Sigma} +
\frac{2(c_I^++\frac{1}{2}c_I^-) }{t+2 \theta} \right) -
Y_{\sss B}(...,-(\Sigma+u))
\frac{-c_I^+ }{t+u+ \Sigma} \right.  \right\}   \nonumber \\ &&
-\sum_{\sss B\{S=-1\}}   Y_{\sss B}(...,t) \left(
\frac{b_I^+}{u-M^2}+\frac{2(b_I^++\frac{1}{2}b_I^-)}{t+2 \theta}
\right) - \sum_{\sss M}\frac{d_I^+}{t-M^2}
W_{\sss B} \left( ...,\frac{-(2u+\Sigma)}{4F} \right) .
\nonumber \\ &&
\label{Psi-B+}
\ee

\be
\varphi_{\sss A^-}(s,t)= &&
\sum_{\sss B\{S=-1\}} b_I^-
\left\{ Y_A(...,t) \left( \frac{1}{s+t+
\Sigma} - \frac{2}{t+2 \theta} \right) -
Y_{\sss A}\bigl( ...,-(s+\Sigma) \bigr)
\frac{1}{s+t+ \Sigma} \right\}     \nonumber \\ && -
\sum_{\sss B\{S=+1\}}c_I^- Y_{\sss A}(...,t) \left(
\frac{1}{s-M^2}+\frac{2}{t+2 \theta} \right)
+\sum_{\sss M} \frac{d_I^-}{t-M^2}
W_{\sss A} \left(...,\frac{2s+\Sigma}{4F} \right)  .
\nonumber \\ &&
\label{varphi-A-}
\ee
%\end{align}
%**************************************
%\begin{align}
\be
\Psi_{\sss A^-}(t,u)  = &&
\sum_{\sss B\{S=+1\}}c_I^- \left\{
Y_{\sss A} \bigl( ...,-(\Sigma+u) \bigr) \frac{1}{t+u+\Sigma}
- Y_{\sss A}(...,t) \left( \frac{1}{t+u+\Sigma} -
\frac{2}{t+2 \theta}\right)  \right\}   \nonumber \\ && +
\sum_{\sss B\{S=-1\}}
b_I^- Y_{\sss A}(...,t)
\left( \frac{1}{u-M^2}+\frac{2}{t+2 \theta} \right) -
\sum_{\sss M} \frac{d_I^-}{t-M^2}
W_{\sss A}\left( ...,\frac{-(2u+\Sigma)}{4F} \right)\, .
\nonumber \\ &&
\label{Psi-A-}
\ee

\be
\varphi_{\sss B^-}(s,t) = && -
\sum_{\sss B\{S=-1\}} \frac{b_I^- }{s+t+ \Sigma}\,
\left\{ Y_B(...,t) - Y_{\sss B} \bigl( ...,-(s+\Sigma) \bigr)
\right\} \nonumber \\ &&
-\sum_{\sss B\{S=+1\}} \frac{c_I^-}{s-M^2}\, Y_{\sss B}(...,t)
+ \sum_{\sss M} \frac{d_I^-}{t-M^2}\,
W_{\sss B} \left( ...,\frac{2s+\Sigma}{4F}\right)\, .
\label{varphi-B-}
\ee
\be
%**************************************
%\begin{align}
\Psi_{\sss B^-}(t,u)= &&
\sum_{\sss B\{S=+1\}} \frac{c_I^-}{t+u+ \Sigma}\,
\left\{ Y_B(...,-(\Sigma+u)) - Y_{\sss B}(...,t) \right\}
\nonumber \\ &&
- \sum_{\sss B\{S=-1\}} \frac{b_I^-}{u-M^2}\, Y_{\sss B}(...,t)
- \sum_{\sss M} \frac{d_I^-}{t-M^2}\,
W_{\sss B}\left( ...,\frac{-(2u+\Sigma)}{4F} \right)\, .
\label{Psi-B-}
%**************************************
\ee

\section{Kaon-Nucleon Couplings to Resonances}
\label{app4}

In this Appendix we give the formula and relations which are
necessary to perform the numerical testing of sum rules.

First of all, we need to know the values of physical triple
couplings
$G_{\overline{K}NR}$
(\ref{Y_X}).
They can be formally obtained from the known decay widths
 $\Gamma_{R \rightarrow \overline{K}N}$.
To connect our
 $G_{\overline{K}NR}$
with
 $\Gamma_{R \rightarrow \overline{K} N}$
listed in
\cite{PDG}
we make use of the standard relation giving particle decay width:
\begin{equation}
\Gamma_{R \rightarrow \overline{K} N}=\frac{1}{8 \pi M^2}
\left.|\vec{k}|
\right|_{CMS} {\sum}'|M_{\alpha \ \  j}^{\; \,  i \beta }|^2.
\label{DecayC}
\end{equation}
The symbol
${\sum}'$
implies the summation over all  allowed final states (isospin and
polarization) and averaging over the initial states of the resonance.

In the case of decay of baryon resonance with the mass parameter
$M$,
strangeness
$S=-1$
and spin
$J=l+\frac{1}{2}$
this sum reads:
\be
%&&
{\sum}'|M_{\alpha \ \  j}^{\; \,  i \beta }|^2
%\nonumber \\ &&
=\frac{1}{2I_R+1}
\sum_{\alpha,i \atop \beta, j}
({P^{u \, (I) }}^{\cdot \, j \, \beta \, \cdot}_{\alpha \, \cdot \; \cdot \,  i})^2
\underbrace{\frac{1}{2l+1} \sum_{j=-l...l \atop \lambda= \pm}
|\mathcal{M}|^2}_{K(l,...)}
%\nonumber \\ &&
\equiv {\mathcal{F}}_{I_R} K(l,M,\mathcal{N})
\nonumber .
\ee
The isotopic factor:
$$
\mathcal{F}_{I_R}=\frac{1}{2I_R+1} \sum_{\alpha,i \atop \beta, j}
( {P^{u \, (I) }}^{\cdot \, j \, \beta \,
\cdot}_{\alpha \, \cdot \; \cdot \,  i})^2=~\left\{ {1, \ \
I_R=0} \atop {1,  \ \ I_R=1} \right. .
$$

With the help of formulae of
Section~\ref{sec-preliminaries}
the decay amplitude for the resonance of normality
$\mathcal{N}= \pm 1$
can be written as follows:
$$
\mathcal{M}=g_R (i)^l k_{\mu_1}...k_{\mu_l} \overline{u}^+(\lambda,p)
\Gamma u^{- \, \mu_1... \mu_l}(j,q).
$$
The matrix
$\Gamma$
is defined by
(\ref{mtrx}).
Calculation of the kinematical factor gives:
$$
K(l,M,\mathcal{N})=|g_R|^2 \frac{l!}{(2l+1)!!} \phi^l [(M
\mathcal{N}-m)^2-\mu^2].
$$
Finally using
(\ref{DecayC}),
(\ref{Y_X})
and
(\ref{phiK})
we  write down the formula expressing the dimensionless
interaction constants
(\ref{dim_constant})
through the corresponding decay widths:
\begin{equation}
G_{ \overline{K}N R}=\frac{8 \pi M^2 \Gamma_{R \rightarrow \overline{K}
N}}{\phi^{\frac{1}{2}}\mathcal{F}_{I_R} [(\mathcal{N}M-m)^2-\mu^2]}.
\label{GKNR}
\end{equation}
The summary of $\overline{K}N R$ couplings employed in our
analysis for the $S=-1$ resonances
obtained from the PDG data on $KN$ spectrum
\cite{PDG}
is presented in
Table~\ref{BaryonDat}.
A formula similar to
(\ref{GKNR})
can be written for the
$KN$
coupling to
$s$-channel
exotic resonances.

\begin{table}[tbp]
\centering
%\begin{ruledtabular}
\begin{tabular}{|lcccl|}
\hline
$\; \; R$ & $I$ & $J$ & $P({\cal N})\ \ $& $G_{ \overline{K}N R}$ \\
%  &  &  &  \\
\hline
$\Lambda(1115)$  &  $0$& $1/2$ &$+(+)$ & $206.\,5 \div 221.\,5 $ \\
$\Lambda(1405)$  & $0$ & $1/2$ & $-(-)$ & $ 6.\,4 \div 6 .\,6$ \\
$\Lambda(1520)$  & $0$ & $3/2$ & $-(+)$ & $ 16.\,2 \div 19 .\,3$ \\
$\Lambda(1600)$  & $0$ & $1/2$ & $+(+)$ & $5.\,7 \div 57.\,0 $\\
$\Lambda(1670)$  & $0$&  $1/2$ &  $-(-)$ & $0.\,15 \div 0.\,35 $\\
$\Lambda(1690)$  & $0$ & $3/2$ & $-(+)$ & $5.\,2 \div 10.\,9 $\\
$\Lambda(1800)$  & $0$ & $1/2$ & $-(-)$ & $ 1.\,1 \div 3.\,5 $\\
$\Lambda(1810)$ &  $0$ & $1/2$ & $+(+)$ & $3.\,1 \div 38.\,7$ \\
$\Lambda(1820)$ &  $0$ & $5/2$ & $ +(+) $& $10.\,0 \div 18.\,1$ \\
$\Lambda(1830)$ &  $0$ & $5/2$ & $-(-)$ & $0.\,04 \div 0.\,23$ \\
$\Lambda(1890)$ &  $0$ & $3/2$ & $ +(-)$ & $0.\,23 \div 1.\,36 $\\
$\Lambda(2100)$ &  $0$ & $7/2$ & $ -(+) $& $2.\,7 \div 13.\,4 $\\
$\Lambda(2110)$ &  $0$ & $5/2$ & $ +(+)$ & $1.\,0 \div 8.\,1 $\\
$\Lambda(2350)$ &  $0$ & $9/2$ & $+(+)$ & $1.\,0 \div 2.\,5$ \\
%\\
\hline
\hline
%\end{tabular}
%\end{ruledtabular}
%\end{table}
%%%%%%%%%%%%%%%%%%%%%%%%%%%%%%%%%%%%%%%%%%%%%%%%%%%%%%%%%%%%%%%%%%%%%%%%%%
%\begin{table}
%\caption{$\Delta$-baryon summary table.}
%\begin{ruledtabular}
%\begin{tabular}{lcccl}
$\; \; R$ & $I$ & $J$ & $P({\cal N})\ \ $& $G_{ \overline{K}N R}$ \\
%  &  &  &  \\
\hline
$\Sigma (1189)$ & $1$ & $1/2$ & $+(+)$ & $141.\,7 \div 144.3 $\\
$\Sigma (1385)$ & $1$ & $3/2$ & $+(-)$ & $0.\,8 \pm 1.\,2 $\\
$\Sigma (1660)$ & $1$ & $1/2$ & $+(+)$ & $2.5 \div 35.1 $\\
$\Sigma (1670)$ & $1$ & $3/2$ & $-(+)$ & $1.\,6 \div 6.\,0 $\\
$\Sigma (1750)$ & $1$ & $1/2$ & $-(-)$ & $0.\,1  \div 1.\,4$ \\
$\Sigma (1775)$ & $1$ & $5/2$ & $-(-)$ & $0.\,85 \div 1.\,27$ \\
$\Sigma (1915)$ & $1$ & $5/2$& $ +(+)$ &$ 0.\,84 \div 5.\,0$ \\
$\Sigma (1940)$ & $1$ & $7/2$ & $+(-)$ & $0.\,44 \div 0.\,79$  \\
$\Sigma (2030)$ & $1$ & $9/2$ & $+(+)$ & $1.\,0 \div 2.\,55 $\\
\hline
\end{tabular}
%\end{ruledtabular}
\caption{\label{BaryonDat} $\Lambda$ and $\Sigma$-baryon summary table. Data
are taken form \cite{PDG}.}
\end{table}

%For our numerical tests we will also need the constants of interaction of
%$\rho(770)$ with kaons and nucleons. These constants can be easily found in
%literature
%(e.g \cite{Nagels}, \cite{EricsonWeise}).
%The only problem is that expressing our constants through that
%defined in literature require an effort.
%$SU(3)$
%prediction for
%$g_{\rho K \overline{K}}$
%is:
%${g_{\rho K \overline{K}}}_{our}=   \frac{1}{2} g_{\rho \pi \pi}$
%where
%$g_{\rho \pi \pi}^2=36.2$.

%Our $\rho$-meson -- nucleon couplings are expressed through vector and
%tensor $\rho-$meson couplings
%$G_V, G_T$
%(see \cite{Nagels} p.14 for the
%definitions)
%in the following way:
%\begin{equation}
%g^{(1)}_{\rho N\overline{N}}=  \frac{G_T}{2m}, \ \ g^{(2)}_{\rho
%N\overline{N}}=  \frac{1}{2} (G^V_\rho-G^T_\rho); \ \   \text{see
%\cite{Nagels}, p. 67:} \ \ \frac{{G^V_\rho}^2}{4 \pi}= 2.78, \
%\frac{G^T_\rho}{G^V_\rho}=6.07. \label{gtKgv}
%\end{equation}
%The final expressions for
%$G_{1,2}^\rho$
%(\ref{Wa})
%give:
%$ G_1^\rho= \pm 15.70 \ \ \ G_2^\rho=\pm 12.32$.
%Note that the sign can not be defined,  however
%our sum rules suggest it should be negative for
%$G_1^\rho$
%and positive for
%$G_2^\rho$.

%%%%%%%%%%%%%%%%%%%%%%%%%%%%%%%%%%%%%%%%%%%%%%%%%%%%%%%%%%%%%%%%%%%%70

\end{document}